\documentclass[journal,12pt,onecolumn,draftclsnofoot]{IEEEtran}
\usepackage{cite}
\usepackage{amsmath,amssymb,amsfonts}
\usepackage{algorithmic}
\usepackage{algorithm}
\usepackage{graphicx}
\usepackage{textcomp}
\usepackage{makecell}
\usepackage{xcolor}
\usepackage{subfigure}
\usepackage[caption=false,font=normalsize,labelfont=sf,textfont=sf]{subfig}
\usepackage{booktabs}

\newcommand{\yh}[1]{\textcolor{orange}{{#1}}}
\newcommand{\hzd}[1]{\textcolor{black}{#1}}
\newcommand{\hzzd}[1]{\textcolor{red}{#1}}
\newcommand{\hdel}[1]{}
\def\BibTeX{{\rm B\kern-.05em{\sc i\kern-.025em b}\kern-.08em
    T\kern-.1667em\lower.7ex\hbox{E}\kern-.125emX}}
\begin{document}
\title{PRINCE: A Pruned AMP Integrated Deep CNN Method for Efficient Channel Estimation of Millimeter-wave and Terahertz Ultra-Massive MIMO Systems \\
}
\author{\IEEEauthorblockN{ Zhengdong Hu, Yuhang Chen, and Chong Han,~\IEEEmembership{Member,~IEEE}}\thanks{Z. Hu, Y. Chen, and C. Han are with the Terahertz Wireless Communications (TWC) Laboratory, Shanghai Jiao Tong
University, Shanghai 200240, China (e-mail: {huzhengdong, yuhang.chen, chong.han}@sjtu.edu.cn).
}}
\maketitle
\thispagestyle{empty}
\pdfoptionpdfminorversion=7
\begin{abstract}
Millimeter-wave (mmWave) and Terahertz (THz)-band communications exploit the abundant bandwidth to fulfill the increasing data rate demands of 6G wireless communications. To compensate for the high propagation loss with reduced hardware costs, ultra-massive multiple-input multiple-output (UM-MIMO) with a hybrid beamforming structure is a promising technology in the mmWave and THz bands. However, channel estimation (CE) is challenging for hybrid UM-MIMO systems, which requires recovering the high-dimensional channels from severely few channel observations. In this paper, a Pruned Approximate Message Passing (AMP) Integrated Deep Convolutional-neural-network (DCNN) CE (PRINCE) method is firstly proposed, which enhances the estimation accuracy of the AMP method by appending a DCNN network. Moreover, by truncating the insignificant feature maps in the convolutional layers of the DCNN network, a pruning method including training with regularization, pruning and refining procedures is developed to reduce the network scale. Simulation results show that the PRINCE achieves a good trade-off between the CE accuracy and significantly low complexity, with normalized-mean-square-error (NMSE) of $-10$ dB at signal-to-noise-ratio (SNR) as $10$~dB after eliminating $80\%$ feature maps.

% \hdel{Specifically, traditional channel estimation (CE) solutions highly rely on prior knowledge of channel statistics, which cannot be accurately captured in practical environment. Moreover, the recently boosted deep-learning (DL) methods suffer from redundant network scale, which is impractical to implement with limited memory and computational capability. To address these problems,}
% \hdel{which allows for low-complexity implementation and is significantly useful to realize efficient CE.} 

\end{abstract}

\section{Introduction}
Millimeter-wave (mmWave) and Terahertz (THz)-band communications explore the potential of ultra-broad bandwidth to meet the demands of high data rates, which are considered as key technologies for 6G wireless communications~\cite{b11,a1}. Nevertheless, the high spreading loss and molecular absorption in mmWave and THz-band severely restrict the communication distance~\cite{b10}. By focusing the transmitted signals to achieve high beamforming gain, ultra-massive multiple-input multiple-output (UM-MIMO) is a promising technology to combat the distance limitation~\cite{b8}. Moreover, hybrid architectures are widely adopted in the UM-MIMO systems to reduce power consumption, which control a large number of antennas with a small number of RF chains~\cite{b9}. To enable the beamforming design of the hybrid UM-MIMO wireless systems, the acquisition of accurate channel state information (CSI) is critical, which strongly depends on reliable and efficient channel estimation (CE).
%\hdel{\yh{you are suggested to change the identification word of each reference. For example, you can use $ref_hybrid_BF_ylf$ or other identification words that are easily to recognize for [5]. Moreover, try to use .bib file to edit the reference, so the order of the reference can be automatically determined.}}
%\yh{Please arrange the references in order, check the entire paper. You can use .bib file as I mentioned above. }

However, CE for the mmWave and THz hybrid UM-MIMO systems is a challenging problem. First, due to the high dimension of antennas with few RF chains in the hybrid architectures, CE requires recovery of the high-dimensional channels based on the received signals severely compressed to the dimension of the number of RF-chains. Second, traditional CE methods cannot achieve satisfactory estimation accuracy without prior knowledge of channel statistics, which is usually unknown and hard to be accurately estimated. Moreover, with the unprecedented large number of antennas in the mmWave and THz UM-MIMO systems, the recently rising deep learning (DL) methods suffer from high computational complexity. 

To this end, the efficient CE scheme with high estimation accuracy for mmWave and THz hybrid UM-MIMO systems is highly in demand.

\subsection{Related Work}
In the literature, CE schemes for MIMO systems mainly include the conventional methods~\cite{b14,b13,a3,a4,a2,b27,new3,b19,new1,new2,new4,new5,new6,new7} and the emerging DL methods~\cite{b30, b17,b15,b21,b22, b24, b6}. On one hand, the conventional methods can be further classified into on-grid and off-grid methods. The on-grid methods treat the angles of arrival and departure as taken from fixed grids, which include the compressive sensing (CS) methods such as the orthogonal matching pursuit (OMP) algorithm~\cite{b14} and the approximate message passing (AMP) algorithm~\cite{b13}. Both OMP and AMP are able to exploit the sparsity of the channel with less pilot training overhead, especially for mmWave and THz channels. However, the performance of these solutions degrades considerably due to the grid-mismatch problem. The grid-mismatch problem is induced by the fact that the parameters in the physical field such as the angles of the propagation path, are continuous and do not fall into the predefined grid precisely. By comparison, the off-grid methods refine the grid resolution or discard the on-grid assumption to improve the accuracy. Particularly, the authors in \cite{a3,a4} iteratively refine the grid to increase the grid resolution. The beamspace 2-D multiple signal classification (MUSIC)~\cite{a2} and dynamic array-of-subarrays (DAoSA)-MUSIC~\cite{b27} methods exploit the subspace by performing eigenvalue decomposition. In \cite{new3}, the direction-of-arrival (DOA) information is estimated by using the 2-D discrete Fourier transform (2D-DFT), and the angle rotation technique is applied to enhance the estimation accuracy. However, these schemes achieve better estimation accuracy than the on-grid methods with the price of high computation complexity.

%\hdel{However, the state-of-art CS methods cannot guarantee high accuracy, owing to the unknown channel sparsity patterns, i.e., the sparse structure of the channel~\cite{b20}.} \hdel{However, the state-of-art CS methods cannot achieve high performance, owning to the unknown sparse structure of the channel, which can not be estimated accurately \cite{b20}.}

On the other hand, with the rapid development of DL tools for wireless communications, many DL-based methods are proposed for CE, which extract the inherent characteristics of the physical channel and can improve the performance of CE~\cite{b30,b17, b15,b21,b22, b24, b6}. The DL-based CE schemes can be further divided into two categories, namely, model-driven~\cite{b30,b17,b15} and data-driven~\cite{b24,b22, b21, b6} methods. The model-driven methods construct the networks with the domain knowledge, usually through the process of unfolding the conventional iterative algorithms. The DL networks only need to learn the parameters required by the iterative algorithm, which are fast and efficient. \hzd{The learned AMP (LAMP) is proposed in \cite{b30} to unfold the AMP iterations into corresponding deep network layers, where the parameters of AMP can be learned and optimized through training with a large dataset. In \cite{b17}, the learned denoising-based approximate message passing (LDAMP) substitutes the original shrinkage function in AMP by a denoising convolutional neural network (DnCNN). The complex-valued Gaussian mixture LAMP (GM-LAMP) based beamspace CE scheme is presented in~\cite{b15} to integrate the LAMP \cite{b30} with a derived Gaussian mixture shrinkage function, which can fully utilizes the prior information of the beamspace channel.} The AMP algorithm is chosen in \hzd{all the aforementioned methods}~\cite{b30, b15, b17} for its powerful capacity in solving the sparse recovery problem and low complexity, \hzd{while the DL network largely enhances the estimation performance of the AMP algorithm}. However, the performance of these model-driven methods is bounded by the underlying AMP algorithm.

By contrast, the data-driven methods~\cite{b21,b22, b24, b6} train the black-box based networks with a large number of data. The authors in \cite{b21} propose a deep neural network (DNN) trained as a black box without relying on the knowledge of channels statics for the doubly selective channels.
The estimator in \cite{b22} employs a customized DNN design. It is based on the deep image prior (DIP) network, which first denoises the received signal, and then applies the least-squares (LS) estimation. \hzd{In \cite{b24}, the DCNN network is trained with a large number of received signals to directly output the important parameters involved in the reconstruction of channels. A spatial-frequency convolutional neural network (SF-CNN) based CE is proposed in \cite{b6} to exploit spatial and frequency correlations, by inputting the channel matrices of adjacent sub-carriers into CNN simultaneously.} Compared to the model-driven methods, the data-driven networks are not bounded by the conventional algorithms and can achieve high accuracy with proper network design and datasets.\hdel{However, these neural networks are considered as black boxes, which may lack guarantee of stability~\cite{b15}. In addition, the size of all the aforementioned networks can easily boom, which makes them not efficient and applicable in practical deployment.}

\hzd{Nevertheless, current design of data-driven DL methods keeps pursuing the high estimation accuracy without considering the applicability in practical deployment. The size of all of the aforementioned networks can easily boom for converging to the best possible performance, which is inefficient and impractical. In addition, the DL network can fall into the trap of over-fitting problem with the increment of network scale, leading that the DL networks lack adaptability to different environments apart from the trained one. As a result, it is essential to explore an efficient DL network with small scale and wide applicability.}

To improve the efficiency of the data-driven DL network, a network slimming method is developed recently~\cite{b3}, by pruning the insignificant output feature maps in convolutional (CV) layers. The network slimming method~\cite{b3} can substantially reduce the model size, run-time memory and computational operations with low network training overhead. Particularly, the feature maps in the CV layers encode the presence or absence, and degree of presence of the features they detect. However, not every feature map can detect the critical features of the network input, leading that there exists feature map redundancy.
On one hand, the network slimming method in~\cite{b3} is useful in pruning the less important feature maps, so that the compact pruned network owns with smaller size. The computational complexity of the network is also reduced by the pruning method. 
On the other hand, since the significant feature maps are maintained after pruning, the attention of DL network is not distracted to unimportant features. This avoids the potential over-fitting problem and improves the robustness of the network. 
To the best of our knowledge, the pruning strategy, which is promising to solve the network redundancy problem, has not been used in the CE DL methods before.

%\yh{I think here, introducing network slimming is enough, there is no need to say CE here}
%\yh{By contrast the relation between and CE is what we proposed, so we can out this in the next subsetion.}
% Related to CE, CV layers are thereby used to excavate the features of the channel structure encoded in the feature maps. However, not every feature map can detect the critical features of channel structure, and hence there exists redundancy of feature maps, which can be eliminated by the network slimming method~\cite{b3}. By pruning the less important feature maps,
% the compact network emerges owning with lower size, reduced computational complexity and still good estimation accuracy. \hzd{Another benefit is that the significant features of the channel can be
% maintained after pruning. As a result, the attention of DL network is protected from the distraction of the pruned unimportant features, which can mitigate the over-fitting problem and improve the robustness.} To the best of our knowledge, the pruning strategy has not been used in the CE DL methods before, which is promising to solve the network redundancy problem.
% \hdel{\yh{You can say the pruning strategy is promising to solve the ... problem in ....}}

\subsection{Contributions}

In this paper, we first propose an AMP integrated DCNN network (AMP-DCNN), which involves with the DCNN network after the AMP algorithm. The CV layers in AMP-DCNN are used to excavate the features of the channel structure encoded in the feature maps, to achieve remarkable performance increment in estimation accuracy. Furthermore, a pruned AMP integrated DCNN CE (PRINCE) method is developed based on AMP-DCNN. By truncating the insignificant feature maps in CV layers, the proposed PRINCE method largely reduces the size of the DCNN network to achieve low complexity, with negligible CE performance degradation. The contributions of this paper are summarized as follows.

\begin{itemize}
    \item \textbf{We propose a novel AMP-DCNN method for CE of the mmWave and THz UM-MIMO systems.} As a generalized example in deploying DL for CE,
    the AMP-DCNN exploits the benefits of traditional AMP CE method and the DL tool. In AMP-DCNN, AMP is first exploited to obtain the coarse CE result. Then, DCNN refines the results from AMP by further enhancing the estimation accuracy with its powerful learning ability.
    
    %Specifically, AMP exploits the channel sparsity and transforms the different dimensions of training pilots into the standard dimension of channel matrix, %which can be easily fed to the DCNN network without processing a large number of unimportant parameters. 
    \item \hzd{\textbf{We propose a PRINCE method by pruning the DCNN network in AMP-DCNN}}. Through training with regularization, pruning and refining, the PRINCE prunes the insignificant feature maps in the CV layers of AMP-DCNN with flexible ratios. The network scale is thereby remarkably decreased in PRINCE, compared to AMP-DCNN, which enables lower computational complexity, while the estimation accuracy reduces negligibly. 
 
    \item \textbf{We conduct extensive experiments to eavalute the CE performance by using practical mmWave and THz UM-MIMO channel datasets.} Results demonstrate that the AMP-DCNN method has noticeably improved estimation accuracy compared to the existing methods. 
    Furthermore, the PRINCE method can effectively truncate $80\%$ DCNN feature maps, reaching a good trade-off between the CE accuracy and low complexity. 
\end{itemize}

The rest of this paper is organized as follows. The system model of the mmWave and THz UM-MIMO systems and the CE problem are investigated in Sec.~\ref{sec:system model}.\hdel{Sec.~\ref{sec:AMP-DCNN} elaborates} The AMP-DCNN method is elaborated in Sec.~\ref{sec:AMP-DCNN}, and the PRINCE method is illustrated in Sec.~\ref{sec: pruned}. Sec.~\ref{sec: evaluation} evaluates the performance of the proposed methods. Finally, the conclusion is drawn in Sec.~VI.

\hdel{\yh{You can ask Prof. Han to check Sec. I, especially Contribution, again. }}
\textbf{Notation:} 
$a$ is a scalar. \textbf{a} denotes a vector. 
\textbf{A} represents a matrix.
$(\cdot)^T$ deﬁnes the transpose.
$(\cdot)^*$ refers to the conjugate transpose.
$E\{\cdot \}$ describes the expectation.
$ \left\|\cdot\right\|_1 $, $ \left\|\cdot\right\|_2 $ and $ \left\|\cdot\right\|_F $ represent the L1 norm, L2 norm and Frobenius norm, respectively.
${\rm vec}\{\cdot \}$ represents the vectorization of the matrix.
$|\cdot|$ denotes the absolute value. $\ast$ represents the convolution operation. $\textbf{I}_N$ defines an $N$ dimensional identity
matrix. $\otimes$ refers to the Kronecker product.

\section{System Model and Problem Formulation}\label{sec:system model}
In this section, we first introduce the system model of the mmWave and THz UM-MIMO systems. Then, we formulate the CE problem. 
\subsection{System Model}

\begin{figure}[t]
    \centering
    \includegraphics[width=1.0\textwidth]{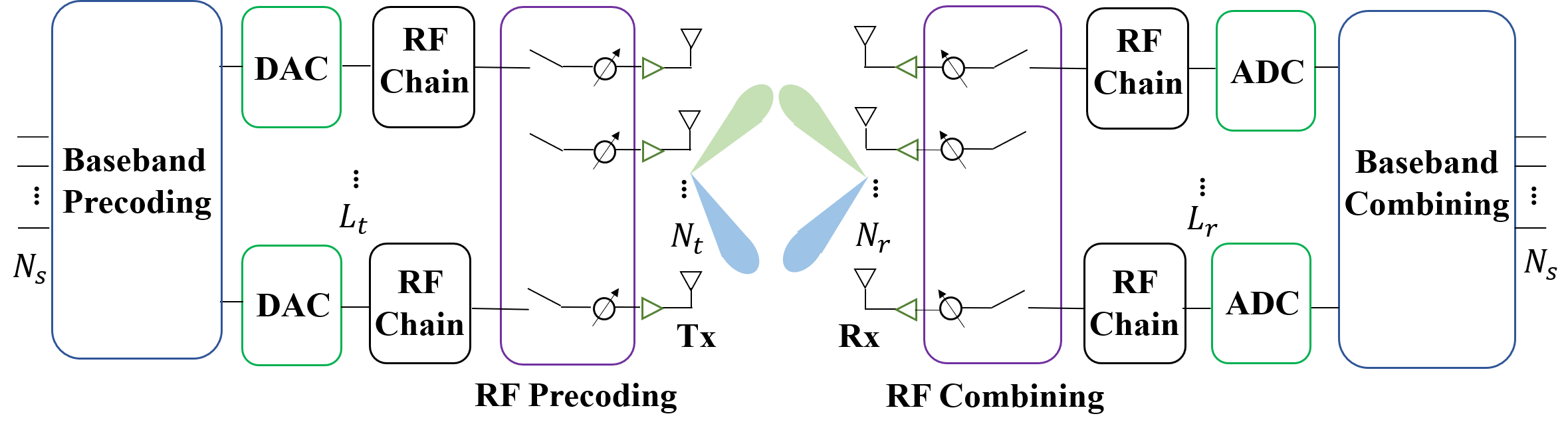} %MIMO3.png
    \caption{Hybrid analog-digital precoding and combining for mmWave and THz UM-MIMO systems.}
    \label{fig}
    % \vspace{-5mm}
\end{figure}

We consider the wideband multi-carrier mmWave and THz UM-MIMO communication systems deploying hybrid precoding and combining architectures at the transmitter (Tx) and receiver (Rx) sides~\cite{b12}, respectively, as shown in Fig.~\ref{fig}. There are totally $K$ subcarriers, with $N_t$ transmit and $N_r$ receive antennas. $L_t$ and $L_r$ RF chains are deployed at Tx and Rx, respectively. 
The systems deploy $N_s$ data streams, satisfying $N_s\leq L_t < N_t$ and $N_s\leq L_r < N_r$ for the considered mmWave and THz hybrid UM-MIMO systems~\cite{b12}. 
At Tx, the transmitted symbol vector $\mathbf{s}[k]\in\mathbb{C} ^{N_s \times 1}$
for the $k^\mathrm{th}$ subcarrier is precoded to obtain the transmitted signal $\textbf{x}[k]\in\mathbb{C} ^{N_t \times 1}$ as
\begin{equation}
    \textbf{x}[k] = \textbf{F}_{\mathrm{RF}}\textbf{F}_{\mathrm{BB}}[k]\textbf{s}[k], \label{e1}
\end{equation}
where $k = 0, \dots, K-1$ indexes the subcarriers. Moreover, $\textbf{F}_{\mathrm{RF}}\in \mathbb{C}^{N_t\times L_t}$ represents the analog precoder, which is implanted by the phase shifters. Therefore, each element in $\textbf{F}_{\mathrm{RF}}$ satisfies the constant module constraint, which can be illustrated as $\textbf{F}_{\mathrm{RF}}[i,j] = \frac{1}{\sqrt{N_t}}{e}^{j f_{i.j}}$, where $f_{i.j}\in [0,2\pi]$ refers to the phase shift value, $i, j$ denote the position of the element in $\textbf{F}_{\mathrm{RF}}$. 
In addition, $\textbf{F}_{\mathrm{BB}}[k]\in \mathbb{C}^{L_t \times N_s}$ represents the baseband digital precoder, which varies for different subcarriers. 

After passing through the channel, the analog and digital combiners $\textbf{W}_{\mathrm{RF}} \in \mathbb{C}^{N_r\times L_r}$ and $\textbf{W}_{\mathrm{BB}}\in \mathbb{C}^{L_r \times N_s}$ are applied to the received signal at Rx, to obtain the baseband received signal $\textbf{y}[k]\in \mathbb{C}^{N_s\times 1}$ as
\begin{equation}
   \textbf{y}[k]=\textbf{W}_{\mathrm{BB}}^*\textbf{W}_{\mathrm{RF}}^*\textbf{H}[k]\textbf{x}[k]+\textbf{W}_{\mathrm{BB}}^*[k]\textbf{W}_{\mathrm{RF}}^*\textbf{n}[k], \label{eq4}
\end{equation}
where $\textbf{H}[k] \in \mathbb{C}^{N_r\times N_t}$ denotes the frequency-domain channel matrix for the $k^\mathrm{th}$ subcarrier, $\textbf{n}[k] \sim \mathcal{CN}(0, \sigma ^2 \textbf{I}_{N_r})$ represents the circularly symmetric complex Gaussian distributed additive noise vector, with noise power equals $ \sigma ^2$. Since the analog combining is implemented by phase shifters, each element in $\textbf{W}_{\mathrm{RF}}$ satisfies the constant module constraint. Furthermore, the digital combiner $\textbf{W}_{\mathrm{BB}}$ is also subcarrier dependent.

We consider a frequency-selective wideband channel model for the mmWave and THz UM-MIMO systems, with a delay tap length $N_c$ in the time domain. Specifically, the channel matrix of the $k^{\rm th}$ subcarrier is described as
\begin{equation}
    \textbf{H}[k]=\sum_{d=0}^{N_c-1}\textbf{H}_de^{-j\frac{2\pi k}{K}d},\label{eq3}
\end{equation}
where $d=0, \dots, N_c-1$ indexes the delay tap. The channel matrix for the $d^\mathrm{th}$ delay tap $\textbf{H}_d \in \mathbb{C}^{N_r\times N_t}$ is represented as
\begin{equation}
    \textbf{H}_d = \sqrt{\frac{N_tN_r}{L}}\sum^L_{l=1}\alpha_lp_{rc}(dT_s-\tau_l)\textbf{a}_R(\theta_{R,l},\phi_{R,l})\textbf{a}_T^*(\theta_{T,l},\phi_{T,l}),
    \label{eq_Hd}
\end{equation}
where $L$ stands for the number of propagation paths, $\alpha_l$ denotes the complex gain of the $l^\mathrm{th}$ path, $p_{rc}(\tau)$ represents the filters including the effects of pulse-shaping and other lowpass filters evaluated at $\tau$. 
Moreover, $T_s$ refers to the sampling period, $\tau_l$ denotes the delay of the $l^\mathrm{th}$ path. The array steering vectors for Rx and Tx are $\textbf{a}_R(\theta_{R,l},\phi_{R,l})\in \mathbb{C}^{N_r\times 1} $ and $\textbf{a}_T(\theta_{T,l},\phi_{T,l})\in \mathbb{C}^{N_t\times 1}$, respectively, in which ($\theta_{R,l},\phi_{R,l}$) and ($\theta_{T,l},\phi_{T,l}$) represent the angle of arrival (AoA) and angle of departure (AOD) pairs. 

We consider uniform planar arrays (UPAs) with $N_x\times N_z$ antennas in the x-z plane. The array steering vector $\textbf{a}(\theta,\phi)$ \hzd{holding for both $\textbf{a}_R(\theta_{R,l},\phi_{R,l}) $ and $\textbf{a}_T(\theta_{T,l},\phi_{T,l})$ }is represented as
\begin{equation}
    \textbf{a}(\theta,\phi) = \frac{1}{\sqrt{N_xN_z}}e^{-j\pi \mathrm{sin}\theta \mathrm{cos}\phi\mathbf{n}_x} \otimes e^{-j\pi \mathrm{sin}\phi\mathbf{n}_z},
\end{equation}
%\begin{equation}
%    \textbf{a}(\theta,\phi) = \frac{1}{\sqrt{N_xN_z}}\sum^{N_x-1}_{n_x= %0}\sum^{N_z-1}_{n_z = 0}e^{jn\pi (n_xsin\theta cos\phi +n_z cos\theta)}
%\end{equation}
where $\mathbf{n}_x=[0,1,\cdots,N_x]^T$ and $\mathbf{n}_z=[0,1,\cdots,N_z]^T$, $N_x$ and $N_z$ denote the number of antennas on x- and z-axis, respectively. 
The channel matrix $\mathbf{H}_d$ in \eqref{eq_Hd} can be represented as a more compact form as 
\begin{equation}
    \mathbf{H}_d=\mathbf{A}_R\mathbf{\Delta}_d\mathbf{A}_T^*,
\end{equation}
where
$\mathbf{A}_R=[\mathbf{a}_R(\theta_{R,1},\phi_{R,1}),\cdots,\mathbf{a}_R(\theta_{R,L},\phi_{R,L})]\in\mathbb{C}^{N_r\times L}$, and $\mathbf{A}_T=[\mathbf{a}_T(\theta_{T,1},\phi_{T,1}),
\cdots, \mathbf{a}_T(\\ \theta_{T,L},\phi_{T,L})]\in\mathbb{C}^{N_t\times L}$, denote array manifold matrices, which contain the array steering vectors for Rx and Tx, respectively.
Moreover, $\mathbf{\Delta}_d=\rm diag(\alpha_1,\cdots, \alpha_L)\in\mathbb{C}^{L\times L}$ is a diagonal matrix containing the path gains. 

The on-grid model approximates the channel using the extended virtual channel model~\cite{b1}.
Specifically, by considering grid of size $G_{rx}\times G_{rz}$ for the AoA and grid of size $G_{tx} \times G_{tz}$ for the AoD, with $G_r = G_{rx}G_{rt}$ and $G_t=G_{tx}G_{tz}$, the channel matrix $\mathbf{H}_d$ can be approximated as
\begin{equation}
    \mathbf{H}_d\approx\tilde{\mathbf{A}}_R\mathbf{\Delta}_d^v\tilde{\mathbf{A}}_T^*,
\end{equation}
where the dictionary matrices
$\tilde{\mathbf{A}}_R=[\mathbf{a}_R(\tilde{\theta}_{R,1},\tilde{\phi}_{R,1}),\cdots,\mathbf{a}_R(\tilde{\theta}_{R,G_r},\tilde{\phi}_{R,G_r})]\in \mathbb{C}^{N_r\times G_r}$ and $\tilde{\mathbf{A}}_T=[\mathbf{a}_T(\tilde{\theta}_{T,1},\tilde{\phi}_{T,1}),\cdots,\mathbf{a}_T(\tilde{\theta}_{T,G_t},\tilde{\phi}_{T,G_t})]\in\mathbb{C}^{N_t\times G_t}$ contain the array steering vectors. Calculation of the grid points at Tx and Rx are similar. Particularly, at Tx, the grid points are obtained as $\mathrm{sin} \tilde{\theta}_{T,g_{tx}}\mathrm{cos} \tilde{\phi}_{R,g_{tx}}\in \{-1 + \frac{2}{G_{tx}},\cdots,1\} $, and $\mathrm{sin} \tilde{\phi}_{T,g_{tz}}\in \{-1 +\frac{2}{G_{tz}},\cdots,1\} $, where $G_{tx}$ and $G_{tz}$ denote the number of grid points of x- and z-axis at Rx, respectively, satisfying $G_{t}=G_{tx}G_{tz}$. Moreover, $g_{tx} = 1,\dots, G_{tx}$, and $g_{tz} = 1,\dots, G_{tz}$. Additionally, $\mathbf{\Delta}_d^v\in \mathbb{C}^{G_r\times G_t}$ forms a sparse matrix with the non-zero elements being the path gains of the quantized spatial frequencies. In this way, $\mathbf{H}[k]$ in~\eqref{eq3} can be expressed as
\begin{subequations}
\begin{align}
    \mathbf{H}[k]&\approx \tilde{\mathbf{A}}_R(\sum_{d=0}^{N_c-1}\mathbf{\Delta}_d^ve^{-j\frac{2\pi k}{K}d})\tilde{\mathbf{A}}_T^*\\ &\approx\tilde{\mathbf{A}}_R\mathbf{\Delta}^v[k]\tilde{\mathbf{A}}_T^*,\label{eqH}
\end{align}
\end{subequations}
% where the channel matrix $\mathbf{H}[k]$ is represented by the sparse matrix
where $\mathbf{\Delta}^v[k]=\sum_{d=0}^{N_c-1}\mathbf{\Delta}_d^ve^{-j\frac{2\pi k}{K}d}\in \mathbb{C}^{G_r\times G_t}$. \hzd{The parameters of the mmWave and THz UM-MIMO systems are summarized in Table~\ref{tab:para}.} 

 \hdel{\yh{We need to explain the meaning of every symbol, here, the meaning of $\tilde{\theta}_{T,l},\tilde{\phi}_{T,l}$ are not explained. You can say, where $\tilde{\theta}_{T,l},\tilde{\phi}_{T,l}$ are grid points of the azimuth and elevation angles, whose values are ... We need to illustrate how exactly these grid points are \textbf{calculated} and selected. 
In addition, these grid points are not related to the number of propagation paths $L$, it is inappropriate to use the subscript $L$. 
During the writing procedure, we want to be careful and critical about the technique we describe, the words and sentences we write, and basic principle we illustrate.
}}

\begin{table}[t] 
\caption{Parameters of the mmWave and THz UM-MIMO systems}
\begin{center}
\begin{tabular}{ccc}
\bottomrule
\textbf{Parameter}& \textbf{Symbol} &\textbf{Unit}\\
\midrule
Number of antennas at Tx & $N_t$ &1\\
Number of antennas at Rx & $N_r$ &1\\
Number of transmitted data streams & $N_s$&1\\
Number of subcarriers & $K$&1\\
Number of RF chains at Tx & $L_t$&1\\
Number of RF chains at Rx & $L_r$&1\\
Transmitted symbol at Tx & $\mathbf{s}$&1\\
Transmitted signal at Tx & $\mathbf{x}$&1\\
Received signal at Rx & $\mathbf{y}$&1\\
Analog precoder at Tx & $\mathbf{F}_{\mathrm{RF}}$ &1\\
Digital precoder at Tx & $\mathbf{F}_{\mathrm{BB}}$&1\\
Training precoder at Tx & $\mathbf{F}_{\mathrm{tr}}$&1\\
Analog combiner at Rx &  $\mathbf{W}_{\mathrm{RF}}$ &1\\
Digital combiner at Rx & $\mathbf{W}_{\mathrm{BB}}$&1\\
Training combiner at Rx &  $\mathbf{W}_{\mathrm{tr}}$ &1\\
Noise vector at Rx & $\mathbf{n}_c$&1\\
Channel matrix & $\mathbf{H}$&1\\
Length of the delay tap length & $N_c$ &1\\
Number of paths & $L$&1\\
Complex gain of the $l^{th}$ path & $\alpha_l$&1\\
Delay of the $l^{th}$ path & $\tau_l$&s\\
Angles pair of departure for the $l^{th}$ path & ($\theta_{T,l},\phi_{T,l}$)&rad\\
Angles pair of arrival for the $l^{th}$ path & ($\theta_{R,l},\phi_{R,l}$)&rad\\
Array steering vector & $\mathbf{a}$&1\\
Number of training frames & $M$&1\\
Measurement matrix &$\boldsymbol{\Phi}$&1\\
Dictionary matrix & $\boldsymbol{\Psi}$&1\\
Vectorized channel vector & $\mathbf{h}$&1\\
\bottomrule
\end{tabular}
\label{tab:para}
\end{center}
% \vspace{-5mm}
\end{table}

\subsection{CE Problem Formulation}
We consider that the channel coherence time is longer than the frame duration, i.e., a static channel is maintained for multiple consecutive frames. During the training process, the pilot signal is transmitted from Tx for $M$ consecutive frames, and the received signals of the $M$ frames are used to reconstruct the channel matrix. For the $m^\mathrm{th}$ training frame at the $k^\mathrm{th}$ subcarrier, the received signal is given by
\begin{equation}
     \textbf{y}^{(m)}[k]=\textbf{W}_{\mathrm{tr}}^{(m)*}\textbf{H}[k]\textbf{F}_{\mathrm{tr}}^{(m)}\textbf{s}^{(m)}[k]+\textbf{n}_c^{(m)}[k], \label{eq5}
\end{equation}
where $m = 1, \dots ,M$ indexes the training frames. The training combiner $\textbf{W}_{\mathrm{tr}}^{(m)}=\textbf{W}_{\mathrm{RF}}^{(m)}\textbf{W}_{\mathrm{BB}}^{(m)}\in \mathbb{C}^{L_r\times N_r}$, while $\textbf{F}_{\mathrm{tr}}^{m}=\textbf{F}_{\mathrm{RF}}^{(m)}\textbf{F}_{\mathrm{BB}}^{(m)}\in \mathbb{C}^{N_t\times L_t}$ denotes the training precoder.
The transmitted pilot symbol $\textbf{s}^{(m)}[k]\in \mathbb{C}^{N_s \times 1}$, with $N_s =L_t$ in the considered system. Moreover, $\textbf{n}^{(m)}_c[k]\in \mathbb{C}^{L_r \times 1} $ refers to the combined noise vector in the $m^{\rm th}$ training frame. 

To reduce the complexity of the system, we consider that $\mathbf{s}^{(m)}[k]$ can be decomposed as $\mathbf{s}^{(m)}[k]=\mathbf{q}^{(m)}{t}^{(m)}[k]$, in which $\mathbf{q}^{(m)} \in \mathbb{C}^{N_s \times 1}$ is independent of frequency. The value of $\mathrm{t}^{(m)}[k]\in \mathbb{C}$ varies among different subcarriers, and is considered to be known at Rx. Moreover, to obtain a subcarrier independent measurement matrix, the transmitted symbol $\mathrm{t}^{(m)}[k]$ can be inverted by multiplying $(\mathrm{t}^{(m)}[k])^{-1}$. As a result, the received signal after the pre-processing is given by
\begin{equation}
     \mathbf{y}'^{(m)}[k]=\mathbf{W}_{\mathrm{tr}}^{(m)*}\mathbf{H}[k]\mathbf{F}_{\mathrm{tr}}^{(m)}\mathbf{q}^{(m)}+\mathbf{n}_c'^{(m)}[k], \label{eq6}
\end{equation}
where $\mathbf{y}'^{(m)}[k]$ and $\mathbf{n}_c'^{(m)}[k]$ are both changed by multiplying $(\mathrm{t}^{(m)}[k])^{-1}$ compared to \eqref{eq5}. After inverting the transmitted symbol $\mathbf{t}^{(m)}[k]$, the received signal can be vectorized to obtain
\begin{equation}
   {\rm vec}\{\mathbf{y}'^{(m)}[k]\}=(\mathbf{q}^{(m)T}\mathbf{F}_{\mathrm{tr}}^{(m)T}\otimes \mathbf{W}_{\mathrm{tr}}^{(m)*}){\rm vec}\{\mathbf{H}[k]\}+\rm vec\{\mathbf{n}_c'^{(m)}[k]\}. \label{eq7}
\end{equation}
The measurement matrix $\boldsymbol{\Phi}^{(m)}$ is defined as $\boldsymbol{\Phi}^{(m)}=(\mathbf{q}^{(m)T}\mathbf{F}_{\mathrm{tr}}^{(m)T}\otimes \mathbf{W}_{\mathrm{tr}}^{(m)*})\in \mathbb{C}^{L_r\times N_tN_r}$. Moreover, according to \eqref{eqH}, the vectorized channel can be represented as $\rm vec\{\textbf{H}[k]\}=(\tilde{\mathbf{A}}_R\otimes\tilde{\mathbf{A}}_T^*)\rm vec\{\mathbf{\Delta}^v[k]\}$. \hzd{The dictionary matrix $\mathbf{\Psi}$ is defined as $\mathbf{\Psi}=(\tilde{\mathbf{A}}_R\otimes\tilde{\mathbf{A}}_T^*)\in \mathbb{C}^{N_rN_t\times G_rG_t}$, and the sparse vector $\mathbf{h}[k]= \rm vec\{\mathbf{\Delta}^v[k]\}\in\mathbb{C}^{ G_rG_t\times 1} $ contains the complex gains of the channel.} Then, \eqref{eq7} can be expressed as 
\begin{equation}
   \rm vec\{ \mathbf{y}'^{(m)}[k]\}=\mathbf{\Phi}^{(m)}\mathbf{\Psi}\mathbf{h}[k]+\rm vec\{\mathbf{n}_c'^{(m)}[k]\}.
    \end{equation}
% \begin{equation}
%     \boldsymbol{\Phi}^{(m)}=(\mathbf{q}^{(m)T}\mathbf{F}_{\mathrm{tr}}^{(m)T}\otimes \mathbf{W}_{\mathrm{tr}}^{(m)*}).
% \end{equation}

During the $M$ training frames, different precoders and combiners are used to compose different $\boldsymbol{\Phi}^{(m)}$. By collecting the received signals as~\eqref{eq7} together, we obtain the received signal model as
\begin{equation}
\underbrace{
\left[\begin{array}{cccc} 
    \rm vec \{\mathbf{y}'^{(1)}[k] \}\\ 
    \vdots\\ 
    \rm vec \{\mathbf{y}'^{(M)}[k]\}
    
\end{array}\right]}_{\textbf{y}[k]}=\underbrace{
\left[\begin{array}{cccc} 
    \boldsymbol{\Phi}^{(1)} \\ 
    \vdots\\ 
    \boldsymbol{\Phi}^{(M)}
    
\end{array}\right]^T}_{\mathbf{\Phi}}\mathbf{\Psi}\mathbf{h}[k]+\underbrace{
\left[\begin{array}{cccc} 
    \rm vec\{\mathbf{n}_c'^{(1)}[k]\} \\ 
    \vdots\\ 
    \rm vec \{\mathbf{n}_c'^{(M)}[k]\}
\end{array}\right]}_{\textbf{n}_c[k]}.\label{eq9}     
\end{equation}
Finally, CE is to solve the sparse reconstruction problem that estimates the sparse vector $\textbf{h}[k]$ 
%\yh{bold} 
in \eqref{eq9}, given by
\begin{align}
   {\rm min}~&\left\|\textbf{h}[k]\right\|_1,  \nonumber \\
   {\rm subject~to}~&\left\|\textbf{y}[k]-\boldsymbol{\Phi}\mathbf{\Psi}\textbf{h}[k]\right\|_2^2 < \epsilon, \label{eq10} 
\end{align}
% \hzzd{where $\epsilon$ is the parameter to represent the maximum error between the measurement $\mathbf{y}[k]$ and the signal $\boldsymbol{\Phi}\mathbf{\Psi}\textbf{h}[k]$.}
where $\epsilon$ is the parameter measuring the estimation error.
\hdel{\yh{what is the meaning of by assuming the estimated channel $\mathbf{h}[k]$ between Tx and Rx. This sentence does not make sense to me. 1. how to assume a channel between Tx and Rx. 2. after (13), you explain that $\mathbf{h}[k]$ is the sparse vector h[k]. But here, you say $\mathbf{h}[k]$ is the estimated channel.
As I said before, we need to make sure the symbols in the paper not be repeatedly defined. I cannot check every detail for you, you need to be responsible for your paper, and check every detail I have mentioned. }}

\hdel{\yh{This sentence is confusing}}
% $\boldsymbol{\Phi}\mathbf{\Psi}\textbf{h}[k]$
\section{AMP-DCNN Method for Channel Estimation}\label{sec:AMP-DCNN}
\begin{figure}[t]
    \centering
    \includegraphics[width=0.9\textwidth]{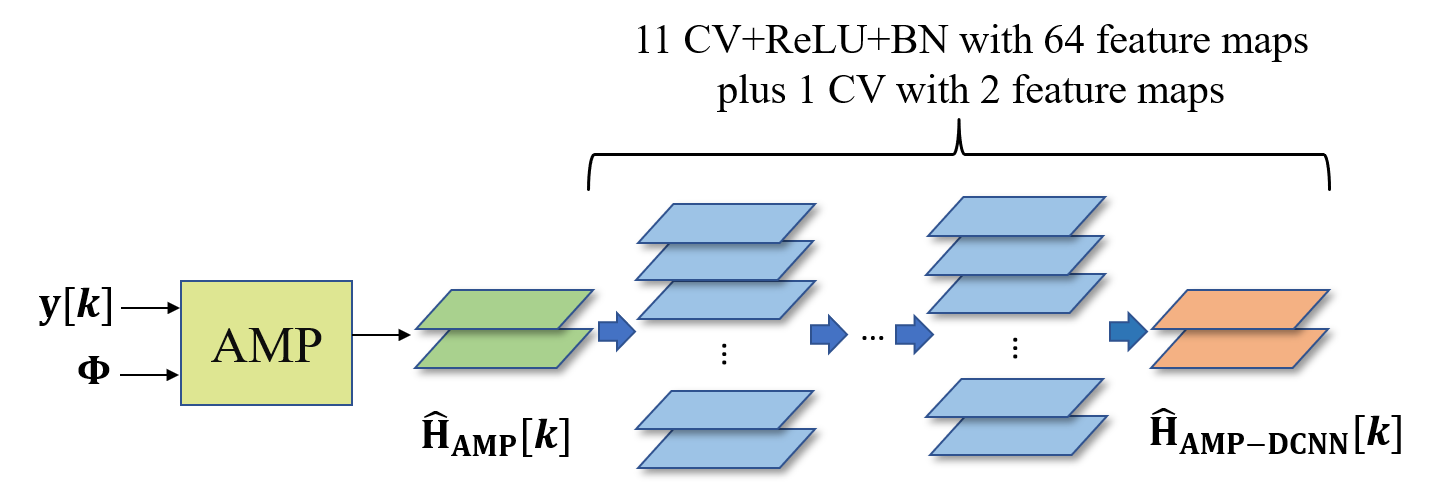}
    \caption{AMP-DCNN based channel estimation.}
    \label{fig11}
\end{figure}

In this section, we propose the AMP-DCNN method to solve the CE problem. The proposed AMP-DCNN is composed of two parts, including the AMP and DCNN, as shown in Fig.~\ref{fig11}. The AMP is used to obtain the coarse channel estimation result from the received signal $\mathbf{y}[k]$ and measurement matrix $\boldsymbol{\Phi}$, while the DCNN refines the output of AMP algorithm to yield the final estimation result with high accuracy.

\subsection{AMP Part}
%To solve the sparse signal recovery problem in \eqref{eq10}, AMP~\cite{b13} is an important iterative CS method.
Due to the huge number of antennas in mmWave and THz UM-MIMO systems, the sparse signal recovery problem in \eqref{eq10} has very high dimension. 
Among the various sparse signal recovery algorithms, the iterative based AMP algorithm is attractive due to its low complexity and fast convergence rate for high-dimensional problems~\cite{b13}.
%For a typical sparse recovery problem to recover N-dimensional vector from $M<N$ observations, AMP simplifies the message passing algorithm by exploiting the large system limit, reducing the propagated messages from $2MN$ to $M+N$. This leads that AMP has low computation complexity especially for high-dimensional sparse signal~\cite{b13}.
%For a typical sparse recovery problem, a N-dimensional vector $\mathbf{h}$ is expected to be recovered from $n<N$ observations through the measurement matrix $\boldsymbol{\Phi}$ by $y=\boldsymbol{\Phi}\mathbf{h}$.
%The computation complexity of AMP is low especially for high-dimensional sparse signal~\cite{b13}, which is commonly used in mmWave and THz UM-MIMO systems.
%while the counterpart OMP algorithm suffers from high complexity when the sparsity level of channel is high~\cite{b15}.
The implementation of the AMP algorithm is illustrated in \textbf{Algorithm 1}. 
Specifically, the received signals $\mathbf{y}[k]$ and the measurement matrix $\boldsymbol{\Phi}$ are input to the algorithm to obtain the estimated result $\hat{\mathbf{H}}_{\mathrm{AMP}}[k]$.

\begin{table}
\centering
\normalsize
\begin{tabular}{l}
\bottomrule
\textbf{Algorithm 1} AMP Algorithm \\
\midrule
\textbf{Input:} Received signal $\mathbf{y}[k]$, measurement matrix $\boldsymbol{\Phi}$,\\~~~~~~~~~number of iterations T.\\
1. Initialization: $\mathbf{r}_{-1}=\mathbf{0}, b_0=0,c_0=0,\mathbf{h}_0=0$\\
2. $(M,N)=\rm size(\mathbf{\Phi})$\\
3. \textbf{for} $t=0,\cdots,T-1$ \textbf{do}\\
4. \hspace{4pt} $\mathbf{r}_t =\mathbf{y}[k]-\boldsymbol{\Phi}\mathbf{h}_t+b_t\mathbf{r}_{t-1}+c_t\mathbf{r}_{t-1}^*$\\
5. \hspace{4pt} $\sigma_t^2=\frac{1}{M}\left\|\mathbf{r}_t\right\|^2_2$\\
6. \hspace{4pt} $\mathbf{z}_t=\mathbf{h}_t+\boldsymbol{\Phi}^T\mathbf{r}_t$\\
7. \hspace{4pt} $\mathbf{h}_{t+1}=\eta_{st}(\mathbf{z}_t;\lambda_t,\sigma_t^2)$\\
8. \hspace{4pt} $b_{t+1}=\frac{1}{M}\sum^N_{i=1}\frac{\partial \eta_{st}(z_{t,i};\lambda_t,\sigma_t^2)}{\partial z_{t,i}}$\\
9. \hspace{4pt} $c_{t+1}=\frac{1}{M}\sum^N_{i=1}\frac{\partial \eta_{st}(z_{t,i};\lambda_t,\sigma_t^2)}{\partial z_{t,i}^*}$\\
10. \textbf{end for}\\
\textbf{Output: $\hat{\mathbf{H}}_{\rm AMP}[k]=\rm reshape(\mathbf{h}_T,[N_r, N_t]$)}.\\
\bottomrule
\end{tabular}
% \vspace{-5mm}
\end{table}

In \textbf{Algorithm 1}, the term $\mathbf{r}_t$ denotes the residual, i.e., the difference between the received signal $\mathbf{y}$ and the recovered signal $\boldsymbol{\Phi}\mathbf{h}_t$. 
The Onsager Correction terms $b_t\mathbf{r}_{t-1}$ and $c_t\mathbf{r}_{t-1}^*$ are used to speed up the convergence of the iterative algorithm. Moreover, $\sigma_t^2$ denotes the estimated noise variance. In each iteration, the estimated result $\mathbf{h}_t$ is updated with the thresholding function $\eta(\mathbf{z}_t;\lambda_t,\sigma_t^2)$, setting the elements less than the threholding level $\lambda_t\sigma_t$ as zero to force the sparsity of the estimated channel $\mathbf{h}_{t+1}$. In addition, $\lambda_t$ is a predefined parameter, while $\sigma_t$ is updated in every iteration. The estimated channel will converge after $T$ iterations, yielding the result of $\mathbf{h}_T$. Finally, $\mathbf{h}_T$ is reshaped into the form of channel matrix $\hat{\mathbf{H}}_{\mathrm{AMP}}[k]$.

However, the estimation result of AMP is not promisingly accurate due to the grid-mismatch problem, i.e., the actual continuous physical parameters do not match the predefined grid of the parameter space. Moreover, the shrinkage parameter $\lambda_t$ takes the same predefined value for all the iterations, which limits the performance of the AMP algorithm. Therefore, to improve the coarse estimated results from the AMP algorithm, we further introduce DCNN to refine the result.

\subsection{DCNN Part}
 The network structure of DCNN network is illustrated in Fig.~\ref{fig11}, where the green and orange parallelograms represent the input and output of the network, respectively. The blue arrows denote the convolution operations. In addition, the blue parallelograms stand for the feature maps in the CV layers. Overall, there are 13 layers in the DCNN architecture, including one input layer, eleven CV layers with 64 filters of the dimensions of 3×3, and one CV layer with 2 filters of the size 3×3, which is indeed the output layer. Moreover, every CV layer except the last one is followed by a batch normalization (BN) layer, and there are eleven BN layers in total.

%Design of each layer in the DCNN is summarized in Table~\ref{tab2}.
To start with, the network input of the DCNN network is constructed with the estimated channel matrix $\hat{\mathbf{H}}_{\mathrm{AMP}}[k]\in \mathbb{C}^{N_r\times N_t}$.
Specifically, $\hat{\mathbf{H}}_{\mathrm{AMP}}[k]$ is separated into the real part $\mathrm{Re}(\hat{\mathbf{H}}_{\mathrm{AMP}}[k])\in \mathbb{N}^{N_r\times N_t}$ and
the imaginary part $\mathrm{Im}(\hat{\textbf{H}}_{\mathrm{AMP}}[k])\in \mathbb{N}^{N_r\times N_t}$ to compose the two channels of the input layer. Followed by the input layer, there are twelve CV layers. Zero padding (ZP) is adopted in each CV layer to keep the dimension of feature map unchanged. Moreover, the rectified linear unit (ReLU) activation function is adopted in each CV layer except the last one, which determines the activated neurons in the network. In particular, the ReLU activation function is represented as 
\begin{equation}
    f_\mathrm{ReLU}(x)=\max(0,x),\label{eq13}
\end{equation}
which is commonly used in the hidden layers for its fast speed of convergence.

In the CV layer, each output feature map can be calculated by convolving its corresponding filter with the previous layer. Specifically, the value of the output neuron in the feature map is obtained by convolving the filter with a small region in the previous layer, which is known as the local receptive region.
The convolution process for each neuron in the CV layer is given by

%\begin{table}[t] 
%\caption{Network Architecture of DCNN}
%\begin{center}
%\begin{tabular}{cc}
%\bottomrule
%\textbf{Layer}& \textbf{Operation} \\
%\midrule 
%1 & Input\\
%2$\thicksim$12 & Conv (64@3$\times$3, ZP, ReLU)+BN\\
%13 & Output (2@3$\times$3, ZP)\\
%\bottomrule
%\hline
%\end{tabular}
%\label{tab2}
%\end{center}
% \vspace{-5mm}
%\end{table}

\begin{equation}\label{eq14}
    h_{{\rm cv}, i,j}^{(m)} = f^{(m)}(\textbf{U}_{{\rm cv},i}^{(m)}\ast \textbf{C}_{{\rm cv},i,j}^{(m-1)}+b_{{\rm cv},i}^{(m)}),
\end{equation}
where $h_{{\rm cv},i,j}^{(m)}$ denotes the value of the $j^\mathrm{th}$ neuron of the $i^\mathrm{th}$ channel in the $m^\mathrm{th}$ CV layer. Moreover, $f^{(m)}(\cdot)$ represents the activation function, which describes the nonlinear mapping relationship. Additionally, $\textbf{U}_{{\rm cv},i}^{(m)}$ and $b_{{\rm cv},i}^{(m)}$ denote the weight and the bias of the $i^\mathrm{th}$ filter, and $\textbf{C}_{{\rm cv},i,j}^{(m-1)}$ stands for the local receptive field for the $j^\mathrm{th}$ neuron of the $i^\mathrm{th}$ channel in the $(m-1)^\mathrm{th}$ layer.

Every CV layer is followed by a BN layer except the output layer, which is used to expedite the convergence of neural networks. The BN performs standardization and normalization operations on the input batch of data, described as 
\begin{subequations}
\begin{align}
   \hat{x} &= \frac{x - \mu_{\textit{B}}}{\sqrt{\sigma_{\textit{B}}^2 + \varepsilon}} \\
   y &= \gamma \hat{x} + \beta \label{eq16}
\end{align}
\end{subequations}
where $x$ and $y$ are the input and output of batch normalization, the subscript $\textit{B}$ represents the current mini-batch, $\mu_{\textit{B}}$ and $\sigma_{\textit{B}}$ denote the mean and standard deviation of the mini-batch of data, $\varepsilon$ is a small number to avoid dividing by zero. Moreover, $\gamma$ and $\beta$ represent trainable parameters to scale and shift the normalized value such that the loss of the neural network is minimized.
 
The last layer of the DCNN network is the output layer, which produces the estimated channel matrix. We do not adopt \hzd{any} activation function in the output layer of the proposed DCNN network, since the activation functions might restrict the range of the output value which is not pre-determined. For example, the output of ReLU activation function in \eqref{eq13} is in the range of $[0,+\infty)$, \hzd{while the values of the elements in the target channel matrix can be less than zero and has no specific bound.}
% In conclusion, the input of DCNN is the estimated channel matrix from the AMP method, while the estimated channel matrix is the output. 
By denoting the input channel matrix as $\hat{\textbf{H}}_{{\rm AMP}}[k]$ and the output matrix as $\hat{\textbf{H}}_{{\rm AMP-DCNN}}[k]$, the end-to-end relationship of the DCNN network is given by
\begin{equation}
    \hat{\textbf{H}}_{{\rm AMP-DCNN}}[k]=f^{(M)}(f^{(M-1)}(\cdot\cdot\cdot f^{(1)}(\hat{\textbf{H}}_{{\rm AMP}}[k]) )),
\end{equation}
where $M$ denotes the number of layers in the DCNN network, $f^{(M)}$ refers to the \hdel{activation function}\hzd{transformation} of the $M^\mathrm{th}$ layer.
During the training process, the mean squared error (MSE) loss function denoted by $l_{\rm MSE}$ is deployed, which is defined as
\begin{equation}
    l_{{\rm MSE}}=\frac{1}{N}\sum_{i=1}^{N}\left\|\textbf{H}^i-\hat{\textbf{H}}^i\right\|_F^2,
\end{equation}
where $N$ denotes the size of the considered dataset of the channel matrix, $\textbf{H}^i$ is the $i^\mathrm{th}$ input channel matrix and $\hat{\textbf{H}}^i$ is the $i^\mathrm{th}$ estimated channel matrix through the DCNN network, respectively. 

Since the DCNN network is designed to minimize the MSE, the resulting number of CV layers increases until the performance of DCNN network saturates, which unfortunately leads that the network involves with a large number feature maps in each layer. To avoid the network redundancy and improve the efficiency and applicability of the DCNN network in practice, we further propose the PRINCE method to prune the DCNN network. 

\section{Pruned AMP Integrated DCNN CE Method}\label{sec: pruned}
In this section, we propose PRINCE method to solve the network redundancy problem. The essence of the PRINCE is to prune the DCNN network in AMP-DCNN by truncating
the insignificant feature maps in CV layers~\cite{b3}. As shown in Fig.~\ref{fig3}, the process of obtaining the pruned DCNN network can be divided into three parts, namely, training with regularization, pruning and refining, respectively. 
\hdel{\yh{I suggest you to re-check this section, try to reduce syntax errors, and make sure each scentence is clear.}}

%\yh{1. $\hat{\mathbf{H}_{\rm AMP}}[k]$. \yh{?}}

%    \caption{Procedure of network slimming method. \yh{This figure is too large. }
%}
%    \label{fig13}
%\end{figure}
%\yh{You can cite the corresponding equations used in this algorithm}\\
%\yh{Mark out the input, output and important intermediate steps as algorithm 1}\\

\subsection{Training with Regularization} 
\hzd{In the DCNN network, CV layers are often used to extract features from data, and encode the features in the feature maps. However, not every feature map can detect the important features. There exist insignificant feature maps in CV layers of the trained AMP-DCNN network, whose output values are close to zero. It is reasonable to prune these feature maps, which contribute little to the final result of DCNN network. However, the number of insignificant feature maps in the DCNN network under normal training is actually small. This means that pruning them brings little compression. Therefore, it is important to create more insignificant feature maps, by condensing more features into a few important feature maps. This can be accomplished by training with regularization.}

Training with regularization prepares more insignificant feature maps by forcing the output of the feature maps close to zero. Specifically, as mentioned in~\eqref{eq16}, each feature map is associated with a scaling factor $\gamma$ in the BN layer, which is multiplied with the output of that feature map. If the scaling factor is close to zero, the output of the corresponding feature map can be scaled close to zero. Training with regularization enables the scaling factors to be near zero by penalizing the value of the scaling factors with regularization. Specifically, the training objective function can be represented as 
\begin{equation}
    L=\underset{(\hat{\textbf{H}}_{{\rm AMP}}[k],\textbf{H}[k])}{\sum}l_{{\rm MSE}}(f(\hat{\textbf{H}}_{{\rm AMP}}[k],\textbf{W}_{\rm DCNN}),\textbf{H}[k])+\lambda \underset{\gamma \in \Gamma }{\sum}g(\gamma),
    \label{eq_loss}
\end{equation}
    where $\hat{\textbf{H}}_{{\rm AMP}}[k]$ and $\textbf{H}[k]$ denote the training input and target output, respectively. In addition, $f$ represents the transformation of the network on the input, $\textbf{W}_{\rm DCNN}$ describes the weights of DCNN network to be trained. Moreover, $l_{{\rm MSE}}$ states the training loss function of the DCNN network, $g(\gamma)=|\gamma|$ is the penalty function to make the value of $\gamma$ close to zero, and $\lambda$ denotes the regularization factor to balance the training loss and penalty term. \hzd{Through training with regularization, more insignificant feature maps arise with the scale factors close to zero.}

% \hzzd{The insignificant feature maps in CV layers can be regarded as the feature maps with output values close to zero, which contribute little to the final result of DCNN network.  As mentioned in~\eqref{eq16}, each feature map is associated with a scaling factor $\gamma$ in the BN layer,} which is multiplied with the output of that feature map.\hdel{If the scaling factor is close to zero, the output of the corresponding feature map after BN layer can be scaled close to zero.} 

% and As a result, the output value of feature maps close to zero can be achieved by enabling the scaling factors to be near zero,

% which is multiplied with the output of that feature map.
% \hdel{If the scaling factor is close to zero, the output of the corresponding feature map after BN layer can be scaled close to zero.} 

\subsection{Pruning} 

% After the network is trained well, the feature maps with the scaling factor close to zero are pruned. This leads that all the incoming and outgoing connections of these feature maps are terminated. The weights and bias of other feature maps still maintain the same after pruning. The process to prune the CV layers in the DCNN network is demonstrated in Fig.~\ref{fig3}.
\begin{figure}[t]
    \centering
    \includegraphics[width=0.80\textwidth]{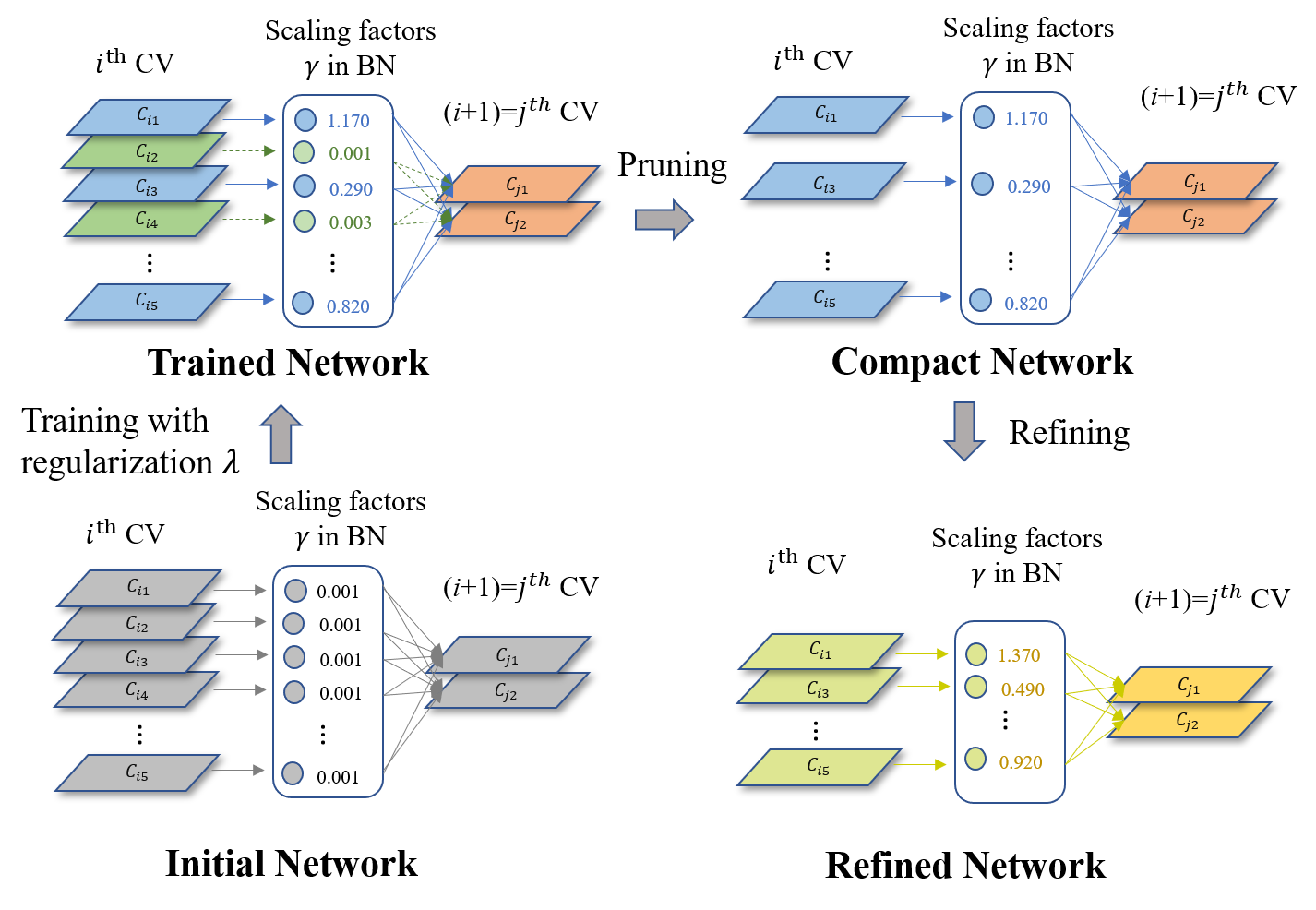}%%Prune-II.PNG v4
    \caption{Process of obtaining the pruned DCNN network.}
\hdel{\yh{Illustration of the network pruning method}}
    \label{fig3}
\end{figure}
After training with regularization, the insignificant feature maps can be pruned to shrink the network scale. The pruning process is demonstrated in Fig.~\ref{fig3}.
In particular, the feature maps with the absolute values of the scaling factors below a certain threshold are pruned. To prune a certain ratio $p$ of feature maps for the network, the threshold value can be determined by collecting the absolute values of the scale factors across all the BN layers into a set. Then, the threshold value is chosen as the value, which is greater than the smallest ratio $p$ of the values in the set and smaller than the rest values. The ratio of pruned feature maps is named as the pruning ratio, denoted by $P_r$. For example, $80\%$ of the feature maps in the CV layers are pruned when the $P_r$ is set as $80\%$. It is worth noticing that the pruning ratio can not be achieved and further increased when all the feature maps in a CV layer are pruned. This is because that the DCNN network would fail with the intermediate CV layer removed. The pruning procedure leads the termination of all incoming and outgoing connections for the pruned feature maps, which reduces the complexity of the network.
In the meantime, the weights and biases of other feature maps still maintain the same after pruning, which keep the important features extracted.

\subsection{Refining} 
Since pruning the feature maps may degrade the estimation performance of the original DCNN network, refining of the pruned network is needed to mitigate this effect. The refining process is mainly to fine-tune the pruned network by training, which has fast convergence. The number of feature maps in the CV layers is substantially decreased after pruning and refining. Meanwhile, the number of computation operations also drops, since the computationally intensive CV layers are pruned. 
As a result, the refined pruned network is slimmer and more efficient with a smaller model size and less computational operations, which can adapt to the limitations of storage and computational resource in different environments. Remarkably, the estimation performance of the pruned network only degrades slightly after the refining process. 
The procedures of the PRINCE are summarized in \textbf{Algorithm 2}.
\hdel{\yh{State here the the procedures of the PRINCS is summarized in Algorithm 2}}

\begin{table}[t] 
\centering
\normalsize
\begin{tabular}{l}
\bottomrule
\textbf{Algorithm 2} PRINCE Algorithm  \\
\midrule
\textbf{Input:} Receive signal $\mathbf{y}[k]$, measurement matrix $\mathbf{\Phi}$, number of iterations T, \\
~~~~~~~~~regularization factor $\lambda$, pruning ratio $p$\\
1. $\hat{\mathbf{H}}_{\rm AMP}[k]=\rm AMP(\mathbf{y}[k],\mathbf{\Phi},T)$ \hdel{\yh{$\hat{\mathbf{H}}_{\rm AMP}[k]$, check the notations}}\\
2. Train the DCNN network by \eqref{eq_loss} with factor $\lambda$\\
3. Prune the trained DCNN network with pruning ratio $p$ \\
4. Refine the pruned DCNN network\\
5. Input $\hat{\mathbf{H}}_{\rm AMP}[k]$ into the refined DCNN to obtain $\hat{\mathbf{H}}_{\rm PRINCE}[k]$\\
\textbf{Output:} $\hat{\mathbf{H}}_{\rm PRINCE}[k]$.\\
\bottomrule
\end{tabular}
\end{table}

%\yh{you can add an algorithm 2 as \textbf{Algorithm 1}, to illustrate the AMP+DCNN+prune method (PRINCE I guess)}

%\begin{figure}[t]
%    \centering
%    \includegraphics[width=0.8\textwidth]{figure/pruned-DCNN-v9.png}%%v3
%    \caption{The DCNN network is slimmer after pruning number of $p_i$ feature maps in the $i^{{\rm th}}$ CV layer for $i=1, \dots ,11$, and resumes the performance after the refining process. \yh{This figure is too large, change the width as [width=0.6 textwidth]. Also, put the figure at the top of each page, use begin{figure}[t]}}
%    \label{fig4}
%\end{figure}

\section{Performance Evaluation and Analysis}\label{sec: evaluation}
In this section, we evaluate the performance of the AMP-DCNN and PRINCE, in terms of convergence, effect of limited training data, estimation accuracy and computational complexity. The estimation accuracy is revealed based on the normalized mean squared error (NMSE), which is defined as 
\newcommand{\norm}[1]{\left\lVert#1\right\rVert}
\begin{equation}
    {\rm NMSE} = \frac{\sum_{k=0}^{K-1}\norm{\textbf{H}[k]-\hat{\textbf{H}}[k]}_F^2}{\sum_{k=0}^{K-1}\norm{\textbf{H}[k]}_F^2}.
\end{equation}

\subsection{Datasets and Simulation Setup}
\begin{figure}
		\centering
		\subfigure[60~GHz channel from the Raymobtime datasets~\cite{b23}.
		%\ch{The figure is not clear. Also, there is no colormap to tell theintensity}
		]{\includegraphics[width=0.45\textwidth,height=6.5cm]{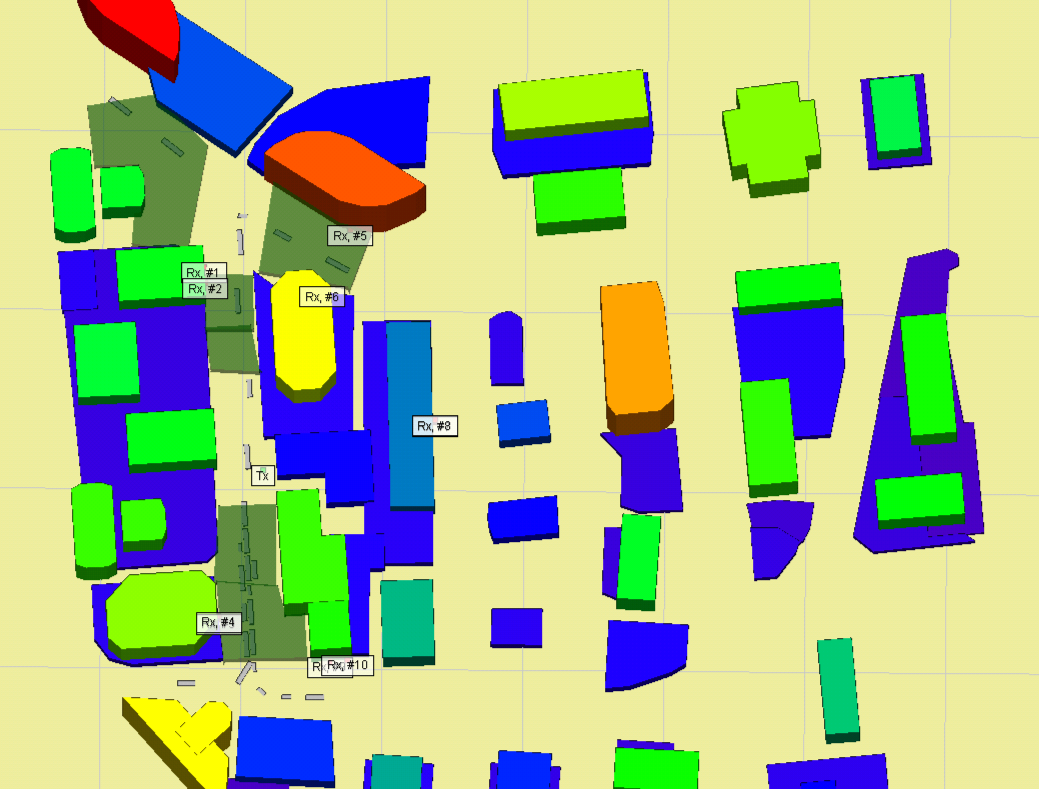}}
		\subfigure[0.3~THz channel generated by Wireless Insite.]{\includegraphics[width=0.45\textwidth,height=6.5cm]{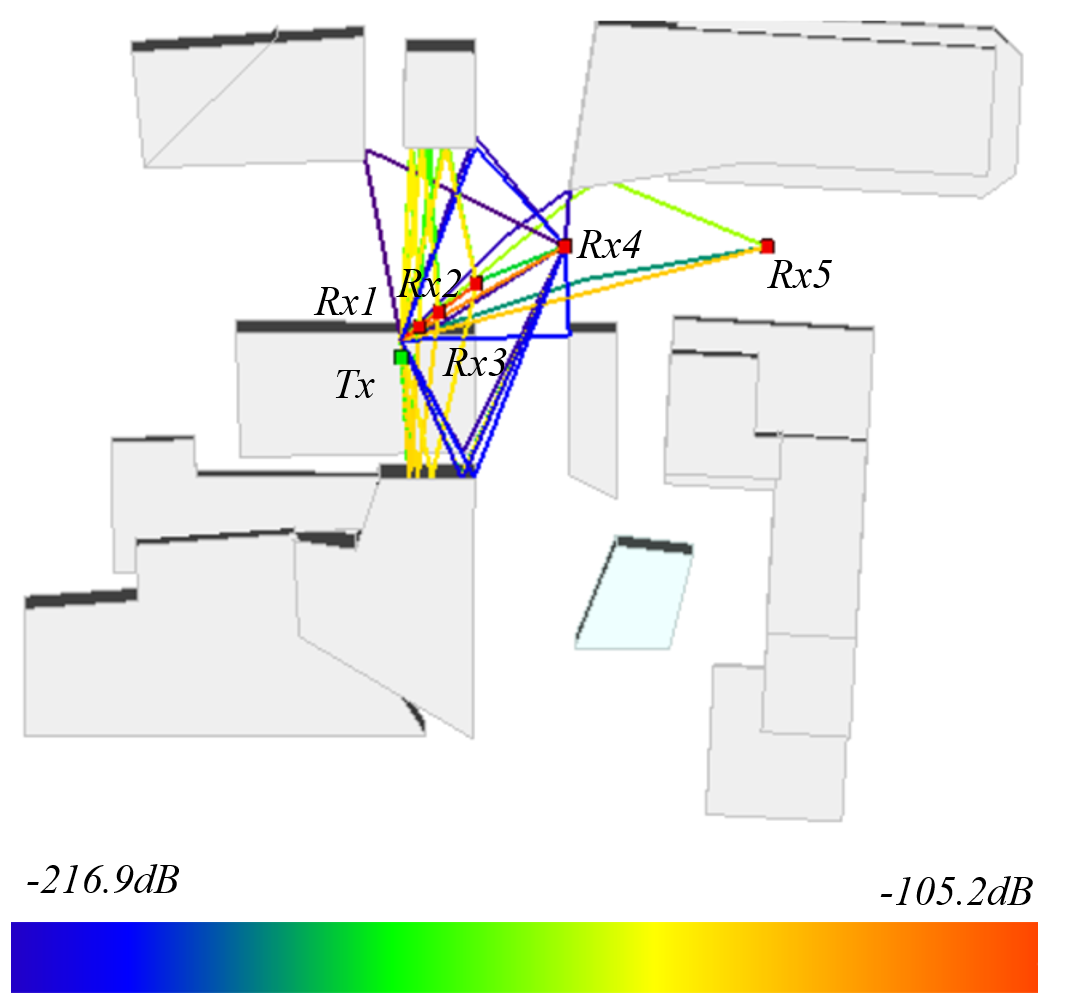}} 
		\caption{Ray-tracing simulation in a 3D scenario. The intensity of the received power of each ray is represented by colors.}%\yh{Fig. 6(a) and (b) can be put in parallel. }} 
		\label{fig_simulation} 
\end{figure}

% We evaluate the performance of the proposed AMP-DCNN method based on two different datasets, including the publicly accessible mmWave channel at 60~GHz \cite{b23}, and self-generated 0.3~THz channel. 
The proposed AMP-DCNN and PRINCE are trained and tested on two different channel datasets generated by the ray-tracing (RT) method, including a publicly accessible mmWave dataset at 60~GHz, and a self-generated THz dataset at 0.3~THz. The mmWave dataset is selected from the Raymobtime datasets~\cite{b23}, which consider 3D scenarios and take the mobility and time evolution of the receivers into consideration. An example of the simulation scenario of the Raymobtime is shown in Fig.~\ref{fig_simulation}(a). Moreover, the Raymobtime datasets incorporate Remcom Wireless Insite~\cite{a5} and a open source Simulator of Urban Mobility (SUMO)~\cite{a6} for mobility simulation. During our evaluation, we select the s002 dataset in the Raymobtime datasets. This dataset is generated at 60~GHz and contains 1000 channels by considering 10 mobile receivers.

\hdel{\yh{ Still need to check the introduction of simulation setup. e.g., how the training dataset is generated, what is the difference between the training dataset and the channel datasets.. How the network is trained after pruning..}}

To obtain the THz dataset, we construct a 3D scenario using Remcom Wireless Insite and measure 1000 channel realizations. 
Specifically, as illustrated in Fig.~\ref{fig_simulation}(b), a typical street with several concrete buildings of different heights and flat terrain are considered in the simulation scenario.
We fix Tx at the top of a building of height 30m, and randomly select 1000 Rx points.
Moreover, we fix the operation frequency at 3.0 THz to obtain 1000 different channel realizations. 
In Fig.~\ref{fig_simulation}(b), we show an example of 5 Rxs, the propagation paths of different path gains are also illustrated. 

% During the experiment, we  Moreover, we extend the datasets at 0.3~THz based on the Wireless Insite simulator.
% As illustrated in Fig.~\ref{fig_simulation}(b), the simulated scenario is a typical street, where several concrete buildings with different
% heights and flat terrain are considered. 1000 Rx points are randomly located to collect the 1000 channel data at the same time.

% An example of the RT simulation in a 3D scenario for the Raymobtime is shown in Fig.~\ref{fig_simulation}(a).

% The Raymobtime dataset incorporates Remcom Wireless Insite and the open source Simulator of Urban Mobility (SUMO) for mobility simulation. 
% During the experiment, we select the 1000 channels of s002 in Raymobtime datasets at 60~GHz with 10 mobile receivers. Moreover, we extend the datasets at 0.3~THz based on the Wireless Insite simulator.
% As illustrated in Fig.~\ref{fig_simulation}(b), the simulated scenario is a typical street, where several concrete buildings with different
% heights and flat terrain are considered. 1000 Rx points are randomly located to collect the 1000 channel data at the same time.
% \begin{figure}
%     \centering
%     \includegraphics[width=0.5\textwidth]{figure/RayMOBTIME.jpg}
%     \caption{ from .  \ch{The figure is not clear. Also, there is no colormap to tell the intensity}}
%     \label{fig8}
% \end{figure}
% \begin{figure}[t]
%     \centering
%     \includegraphics[width=0.4\textwidth]{figure/map.png}
%     \caption{Simulation environment for the THz channel dataset.}
%     \label{fig14}
%     \vspace{-5mm}
% \end{figure}

The simulation parameters are selected as follows. For the mmWave channel, the number of antennas and RF chains at Tx and Rx are $N_t = 64$, $L_t = 2$, $N_r = 16$ and $L_r = 4$, respectively.\hdel{leading that the dimension of the training input is $64 \times 16 \times 2$.} Moreover, the number of subcarriers and pilots are $K=256$ and $M=100$.
The corresponding parameters for the THz channel are $N_t = 512$, $L_t=4$, $N_r = 32$, $L_r=8$, $K=16$ and $M=1000$, respectively.\hdel{while the dimension of the \hzzd{training input to the DCNN network} is $512\times 32 \times 2$.} The received signal $\mathbf{y}[k]$ for both mmWave and THz frequencies is obtained as in \eqref{eq9}, in which the measurement matrix $\mathbf{\Phi}$ is the same for every channel $\mathbf{h}[k]$. To generate the measurement matrix $\mathbf{\Phi}$, the symbol $\mathbf{q}^{(m)}$ is generated according to the normal distribution, while the precoder $\mathbf{F}_{tr}^{(m)}$ and the combiner $\mathbf{W}_{tr}^{(m)}$ are generated with the phase shift values uniformly distributed. In terms of noise, 6 different values of SNR from $-5$ dB to 20 dB are considered. Next, the received signals are pre-processed by AMP to generate the coarse estimated channel matrices. Followed by that, the estimated channel matrices at different SNRs are mixed together, which compose the dataset to train the DCNN network. Therefore, both mmWave and THz datasets contain 6000 elements. After generating the datasets, the DCNN network is trained with the adaptive moment estimation (Adam) optimizer for its fast convergence rate, and the learning rate is set as $l_r = 10^{-3}$. Finally, the PRINCE method is operated with the regularization factor $\lambda=10^{-4}$, which is empirically chosen from the values $10^{-3}$, $10^{-4}$ and $10^{-5}$. Moreover, the pruning ratios for PRINCE are selected as values from $10\%$ to $70\%$ and from $10\%$ to $80\%$ for mmWave dataset and THz dataset, respectively.

% Then, the mmWave and THz channel datasets are constructed to generate the pilot matrix $\boldsymbol{\Phi}$ and received signal matrix $\textbf{Y}$ in~\eqref{eq9} under 6 different values of SNR from $-5$ dB to 20 dB. Next, the pilots and received signals are passed to AMP to generate the estimated channel matrix. After AMP,  the estimated channel matrix at different SNRs are mixed together as the dataset to train the DCNN network, which has the data size of 6000 for the mmWave and THz channel datasets, respectively.

All the experimental results are implemented on a PC with Intel(R) Xeon(R) CPU E5-2690 v4 @ 2.60 GHz and an Nvidia GeForce RTX 2080 Ti GPU. In addition, the simulation of the UM-MIMO system and the conventional CE nethods including OMP~\cite{b14}, AMP~\cite{b13} and MMSE~\cite{b19}, are operated in Matlab (R2018a) environment, while SF-CNN~\cite{b6}, the proposed DCNN network and the pruning method are carried out in the Visual Studio framework.

\subsection{Convergence Evaluation}
\hdel{\yh{In the simulation part, try to describe each experimental result with 3 sentences. 1. summarize the observed phenomenon. 2. give detailed example. 3. explain the reasons. Please check the entire section. }}
\begin{figure}[t]
    \centering
    \includegraphics[width=0.65\textwidth]{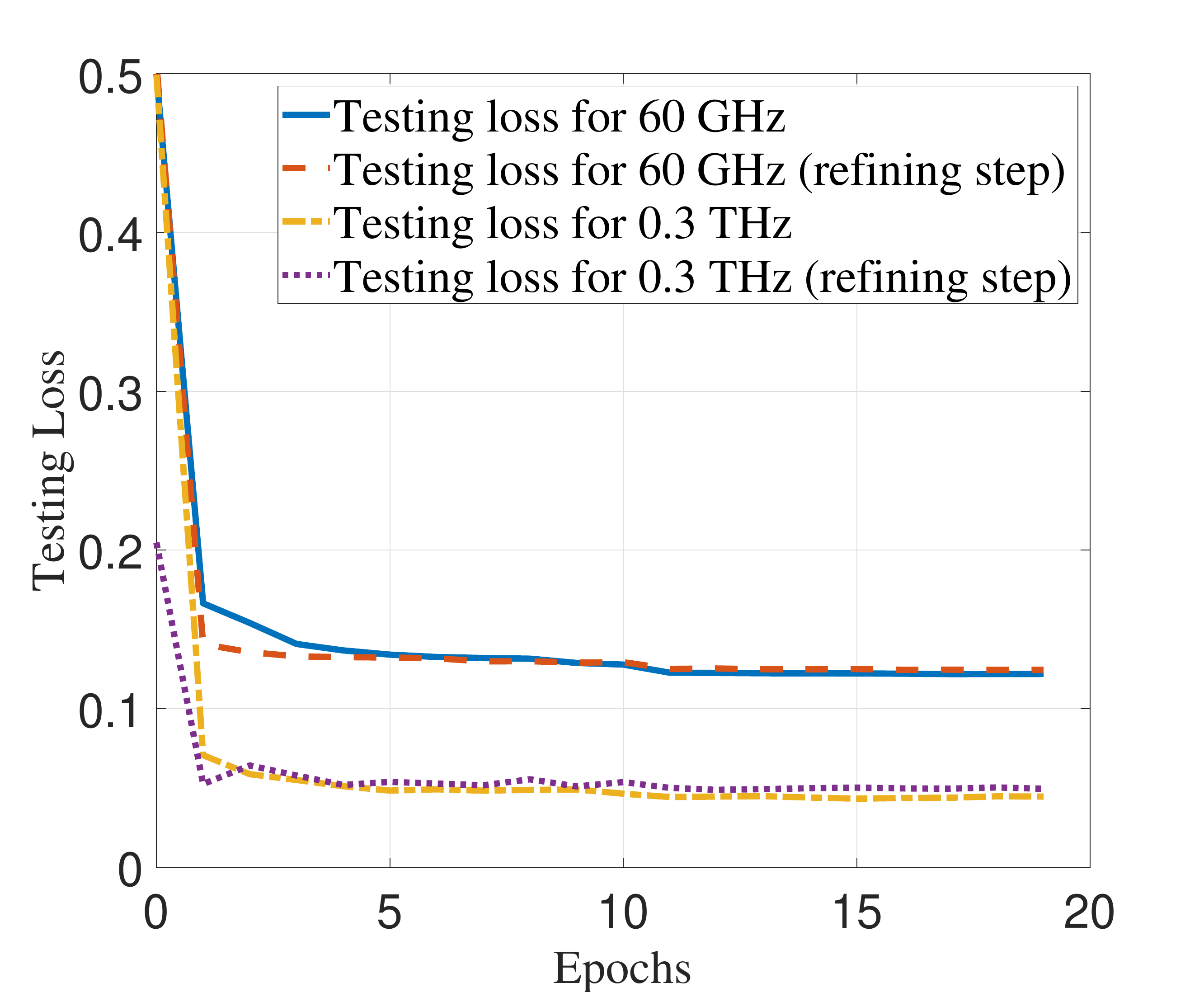}
    \caption{Testing loss of the DCNN network for 60 GHz and 0.3 THz. }
    \label{fig_convergence}
\end{figure}
The convergence performance of the proposed DCNN model for both mmWave and THz channels is evaluated in Fig.~\ref{fig_convergence} by plotting the testing loss versus the number of epochs. 
It is observed that the testing losses are stable after 10 epochs for the two scenarios, which verifies the fast convergence performance of the DCNN model.\hdel{Specifically, after the $10^{\rm th}$ epoch, the testing loss for 60 GHz keeps constant at the value of 0.12, and the testing loss for 0.3 THz stays at the value of 0.043.} By contrast, the refining step after pruning 10$\%$ feature maps takes less than 5 epochs to converge, which incurs little training overhead for PRINCE and still yields expedite convergence. The fast convergence is because that the weights and biases of the unpruned feature maps maintain the same after pruning, and only minor changes to these parameters in the refining step are required to converge.

\hdel{\yh{Specifically, ...., gave detailed example with number. You need to describe the changes of the lines, and explain the reasons causing these changes.}}

\subsection{Effect of Limited Training Data}
\begin{figure}[t]
    \centering
    \includegraphics[width=0.65\textwidth]{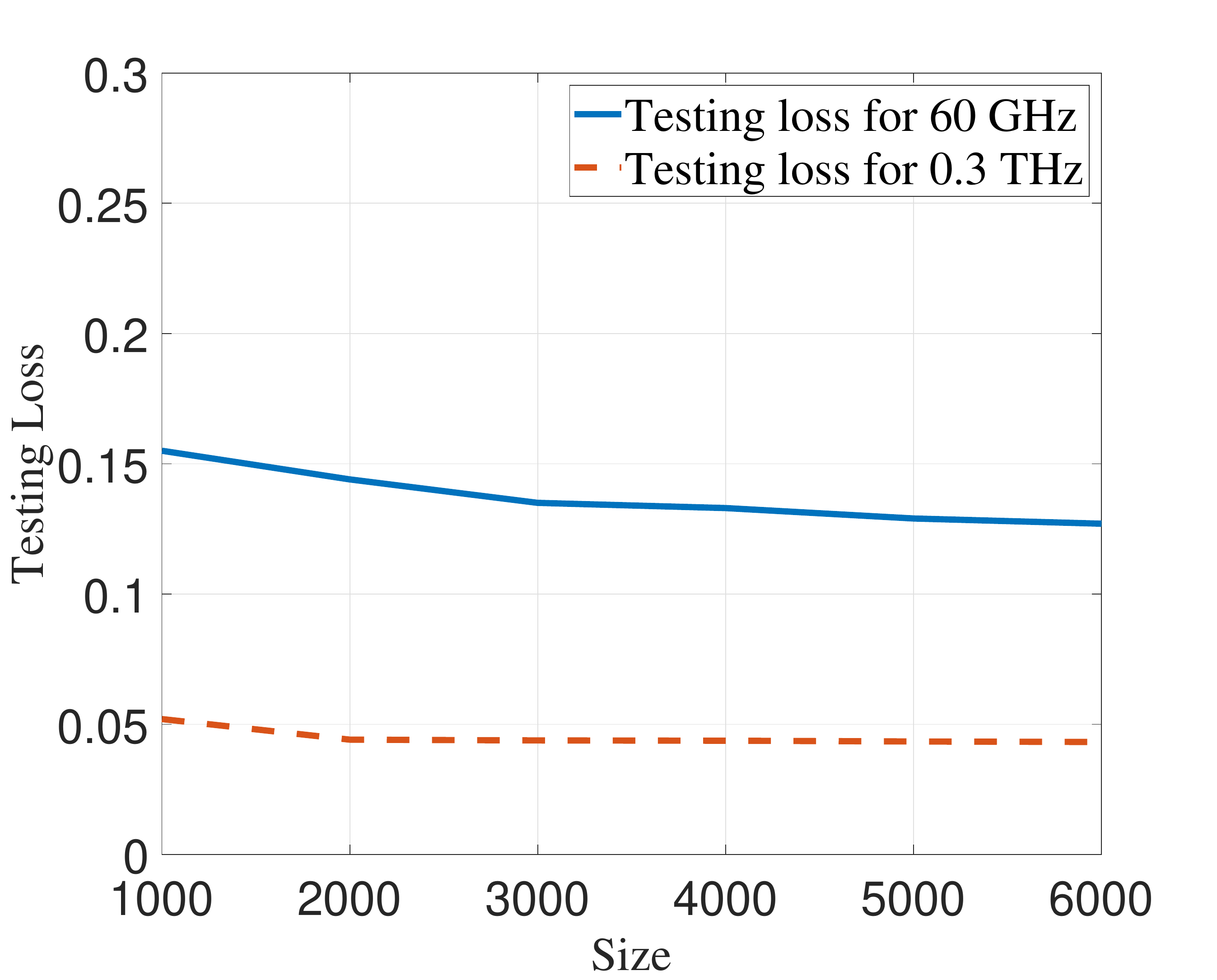}
    \caption{Testing loss with different sizes of training data set.}
    \label{fig_stable}
\end{figure}

In the practical deployment of CE schemes, DL methods are usually faced with the problem of limited training data set due to the limitation of time and resources, and it is essential for the DL method to maintain high estimation accuracy with a limited training dataset. To evaluate the performance of the proposed DCNN network with limited training resources, we train the proposed DCNN network under different sizes of training sets for 20 epochs. 
As shown in Fig.~\ref{fig_stable}, the proposed DCNN network still has the ability to converge even with only 1000 samples in the training set.
%\yh{example and reason}
Moreover, the testing loss decreases with the size of the training set, which indicates that the performance of the DCNN network improves with the increasing of data. 
%\yh{example and reason}
In addition, when the size of the dataset decreases by $83\%$ from 6000 to 1000, the testing loss only increases by $28\%$ for 60~GHz and $17\%$ for 0.3~THz channels, respectively. 
Therefore, there is no drastic increase of the testing loss as the size of dataset decreases. We can thereby state that the performance of the proposed DCNN network is robust with limited training data.
\subsection{Estimation Accuracy}
To assess the estimation accuracy of the proposed AMP-DCNN model, the CS-based methods including the OMP~\cite{b14}, AMP~\cite{b13}, the conventional CE schemes of non-ideal MMSE estimator~\cite{b19} and the data-based DL method SF-CNN~\cite{b6} are included for comparison. In addition, to show the advantage of AMP algorithm against other CS-based methods in the AMP-DCNN method, OMP-DCNN is constructed for comparison by replacing the AMP algorithm with the OMP algorithm.
For fair comparison, the same datasets including the mmWave channel and the THz channel are used in these methods.
%\subsubsection{\hzd{LAMP}}
%\hzd{LAMP~\cite{b30} unfolds the iteration of AMP algorithm into the DL network, which learns the parameters in every iteration through training in a large dataset. Compared to AMP algorithm, LAMP optimizes the intermediate parameters to approach the ideal performance of AMP. However, since the essence of LAMP is still the same as AMP method to solve the sparse recovery problem, the performance of LAMP is bounded by the original AMP algorithm.}
%\subsubsection{\hzd{SF-CNN}}
%\hzd{SF-CNN~\cite{b6} based CE exploits the spatial and frequency correlation by inputting the coarse estimation of channel matrices at adjacent subcarriers into the CNN simultaneously. Instead of applying the AMP method to recover the channel matrices from the measurement data, SF-CNN uses a tentative estimation module. The tentative estimation just multiplies the received signal matrix $\mathbf{Y}$ by the training combiner $\mathbf{W}_{tr}$ and the transpose of training precoder $\mathbf{F}_{tr}$, defined as $\hat{\mathbf{H}}=\mathbf{W}_{tr}\mathbf{YF}_{tr}^{H}$. To be compared with the AMP-DCNN method, the number of CV layers and feature maps are kept the same for these two data-based methods.}

%\subsubsection{\hzd{OMP-DCNN}}
%\hzd{OMP-DCNN is the extension of AMP-DCNN by replacing the AMP algorithm by the OMP algorithm. The performance of OMP-DCNN method is compared with AMP-DCNN method to show the advantage of the AMP algorithm against other CS-based methods for the coarse estimation of channel matrix before the DCNN network.}

\begin{figure}[t]
    \centering
    \subfigure[60 GHz channel.]{\includegraphics[width=0.48\textwidth]{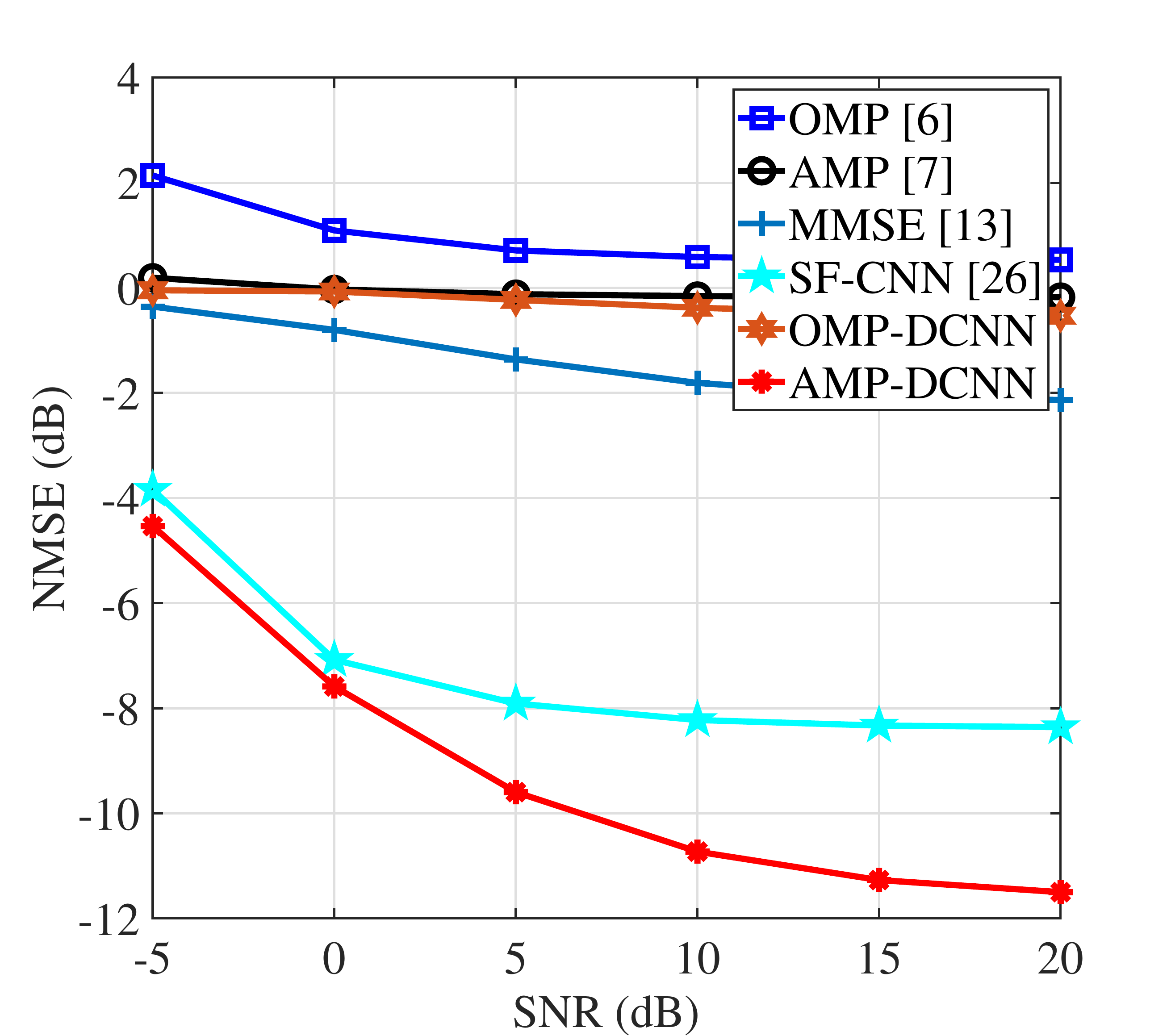}} %comp2.eps
    \subfigure[0.3 THz channel.]{ \includegraphics[width = 0.48\textwidth]{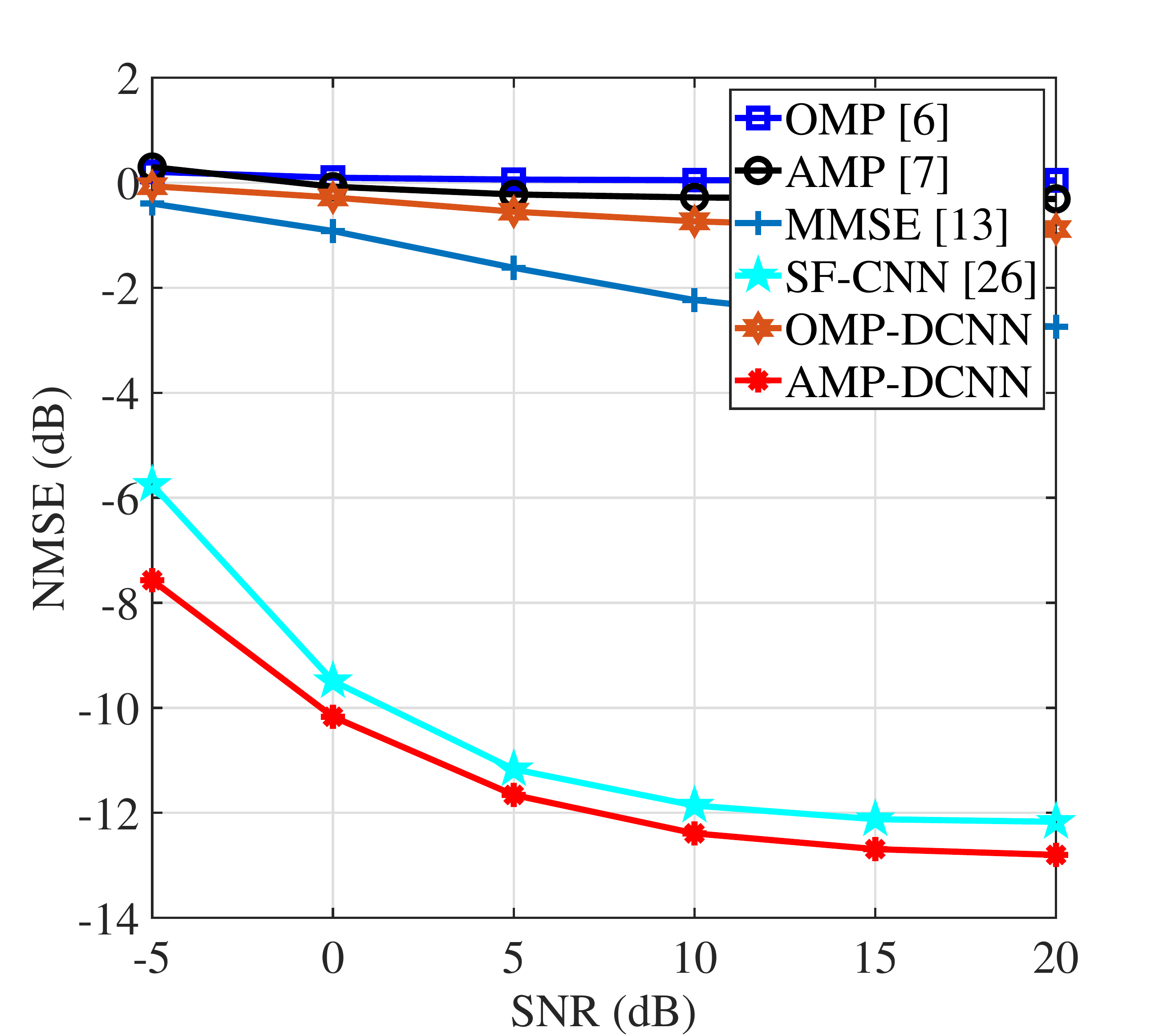}}
    \caption{Channel estimation accuracy \hzd{of the AMP-DCNN algorithm.}\hdel{\yh{the line width of of the frame should be the same.}}}
    \label{fig6}
    % \vspace{-5mm}
\end{figure}

As shown in Fig.~\ref{fig6}(a), based on the mmWave channel dataset, the proposed AMP-DCNN method achieves the lowest NMSE among the six different methods, in which the NMSE of the proposed AMP-DCNN method reaches $-10$~dB when SNR = 10~dB, while the NMSE values of the other five methods are above $-10$~dB for the six different SNR values. Furthermore, based on the THz channel dataset, the proposed AMP-DCNN method outperforms the benchmark methods as shown in Fig.~\ref{fig6}(b). 
The NMSE is lower than $-10$~dB at $5$~dB SNR for the AMP-DCNN method. No matter which dataset the estimation is evaluated on, AMP-DCNN has shown high accuracy in CE.

\begin{figure}[t]
    \centering
    \subfigure[60 GHz channel.]{\includegraphics[width=0.48\textwidth]{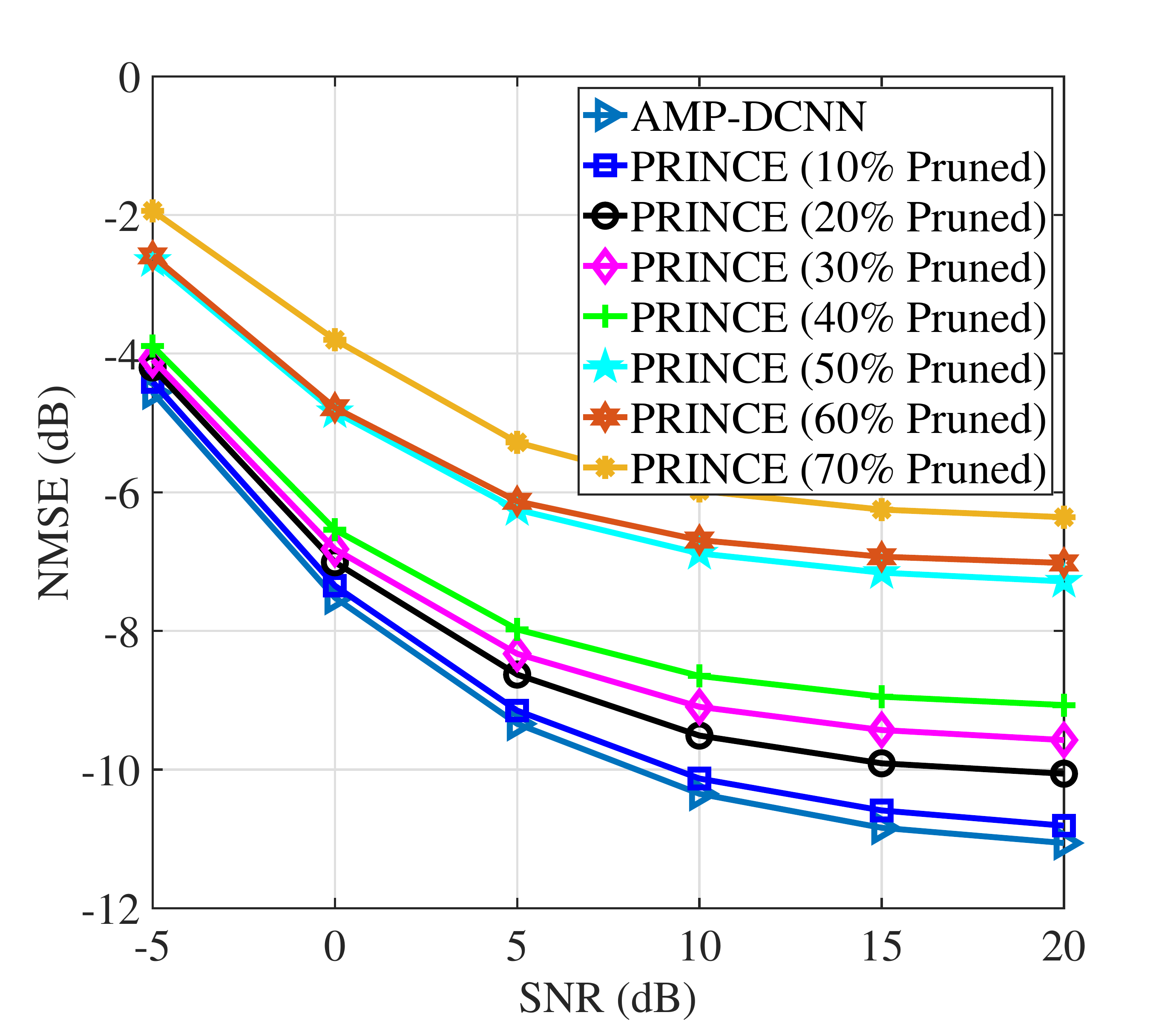}}%%mm_prune
    \subfigure[0.3 THz channel.]{\includegraphics[width=0.48\textwidth]{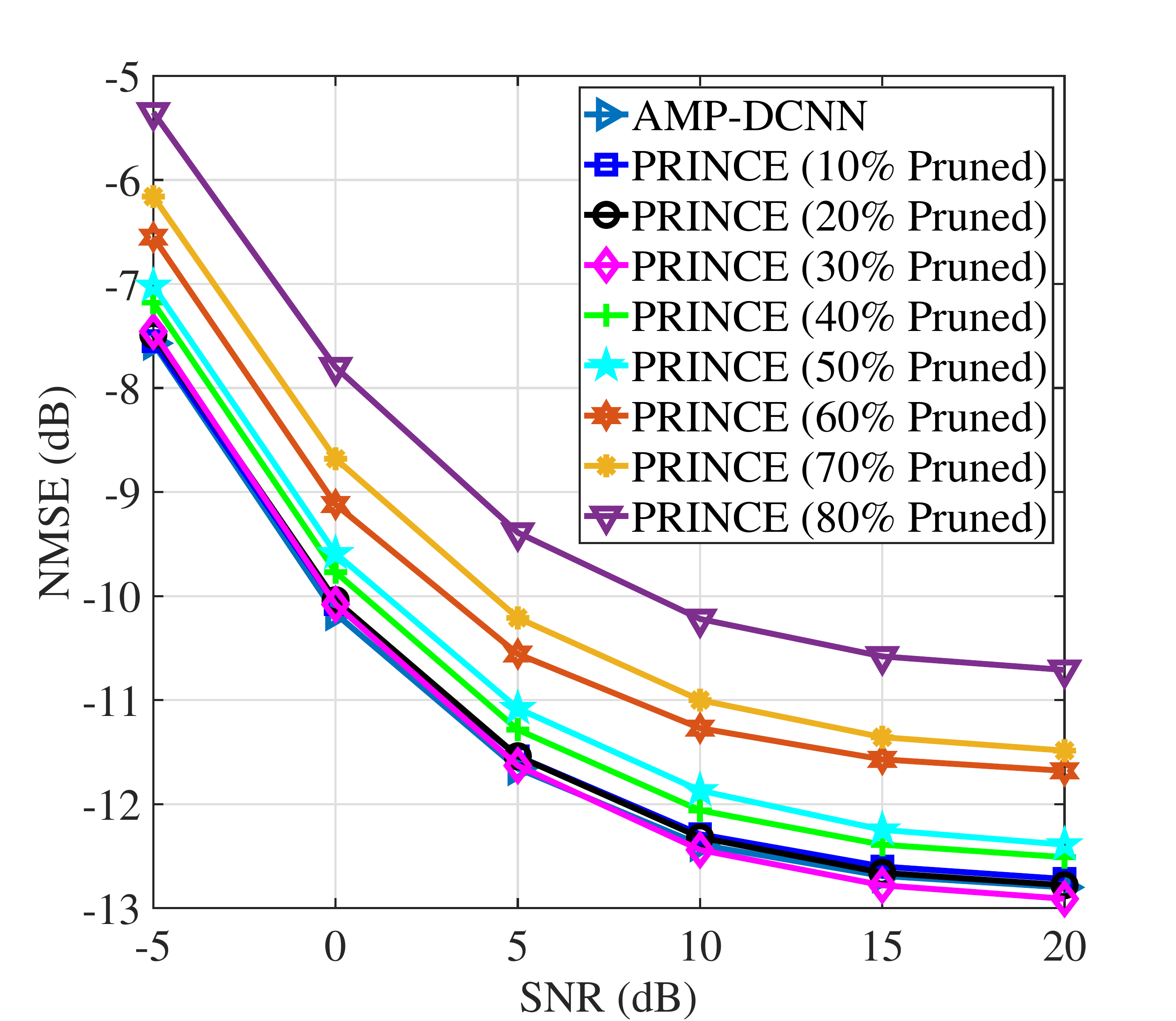}}%%thz_prune
    \caption{Channel estimation accuracy of the pruned AMP-DCNN algorithm. }
    \label{fig7}
    % \vspace{-5mm}
\end{figure}

Moreover, we vary the pruning ratio, and analyze the estimation accuracy of the PRINCE algorithm, upper-bounded by the AMP-DCNN method. In Fig.~\ref{fig7}(a), based on the mmWave channel, we find that the estimation accuracy degrades by less than $1\%$ and $15\%$, when the pruning ratio increases from $10\%$ to $40\%$. At 10~dB SNR, the NMSE lies below $-6$~dB for the PRINCE with the pruning ratio at $70\%$. In addition, the performance of PRINCE based on the THz channel with different pruning ratios is shown in Fig.~\ref{fig7}(b). The performance of the network pruning thirty percent is nearly the same as the uncontacted AMP-DCNN network. There is roughly 2~dB degradation on average for different SNR values, after pruning eighty percent of feature maps of the DCNN.

Overall, AMP-DCNN has advantages over the conventional methods in terms of the estimation accuracy in both mmWave and THz channel datasets, which presents the superiority in accurate CE and great adaptability in different scenarios.
More importantly, the pruning of insignificant channels compromises the accuracy reasonably, which makes PRINCE method appealing for practical implementation. 

\hdel{\subsection{\hzd{Robustness}}
\hzd{In this section, to embody the gain of robustness from AMP method and pruning method,  the robustness of SFCNN, AMP-DCNN and pruned AMP-DCNN methods in two different environments is compared with each other.} }

\subsection{Computational Complexity}

\begin{table}[t]
\caption{Comparison on computational complexity.}
\begin{center}
\begin{tabular}{cc}
\bottomrule
\textbf{Method of CE}& \textbf{Complexity} \\
\midrule
OMP~\cite{b14} & $\mathcal{O}(SN_tN_r)$\\
AMP~\cite{b13}& $\mathcal{O}(TN_tN_r)$\\
MMSE~\cite{b19} & $\mathcal{O}(N_t^3N_r^3)$\\
%\hzd{LAMP~\cite{b30}} & \hzd{$\mathcal{O}(TN_tN_r)$}\\
\hzd{SF-CNN~\cite{b6}} & \hzd{$\mathcal{O}(N_tN_r((N_t+N_r)+\sum_{l=1}^{l_c}F_l^2N_{l-1}N_{l}))$}\\
\hzd{OMP-DCNN} & \hzd{$\mathcal{O}(N_tN_r(S+\sum_{l=1}^{l_c}F_l^2N_{l-1}N_{l}))$}\\
AMP-DCNN & \makecell[c]{
$\mathcal{O}(N_tN_r(T+\sum_{l=1}^{l_c}F_l^2N_{l-1}N_{l}))$}\\
PRINCE & \makecell[c]{$\mathcal{O}(N_tN_r(T+
\sum_{l=1}^{l_c}F_l^2(N_{l-1}-p_{l-1})(N_{l}-p_l)))$}\\
\bottomrule
\end{tabular}
\label{tab1}
\end{center}
% \vspace{-5mm}
\end{table}

\begin{figure}[t]
    \centering
    \subfigure[60 GHz channel.]{\includegraphics[width=0.45\textwidth]{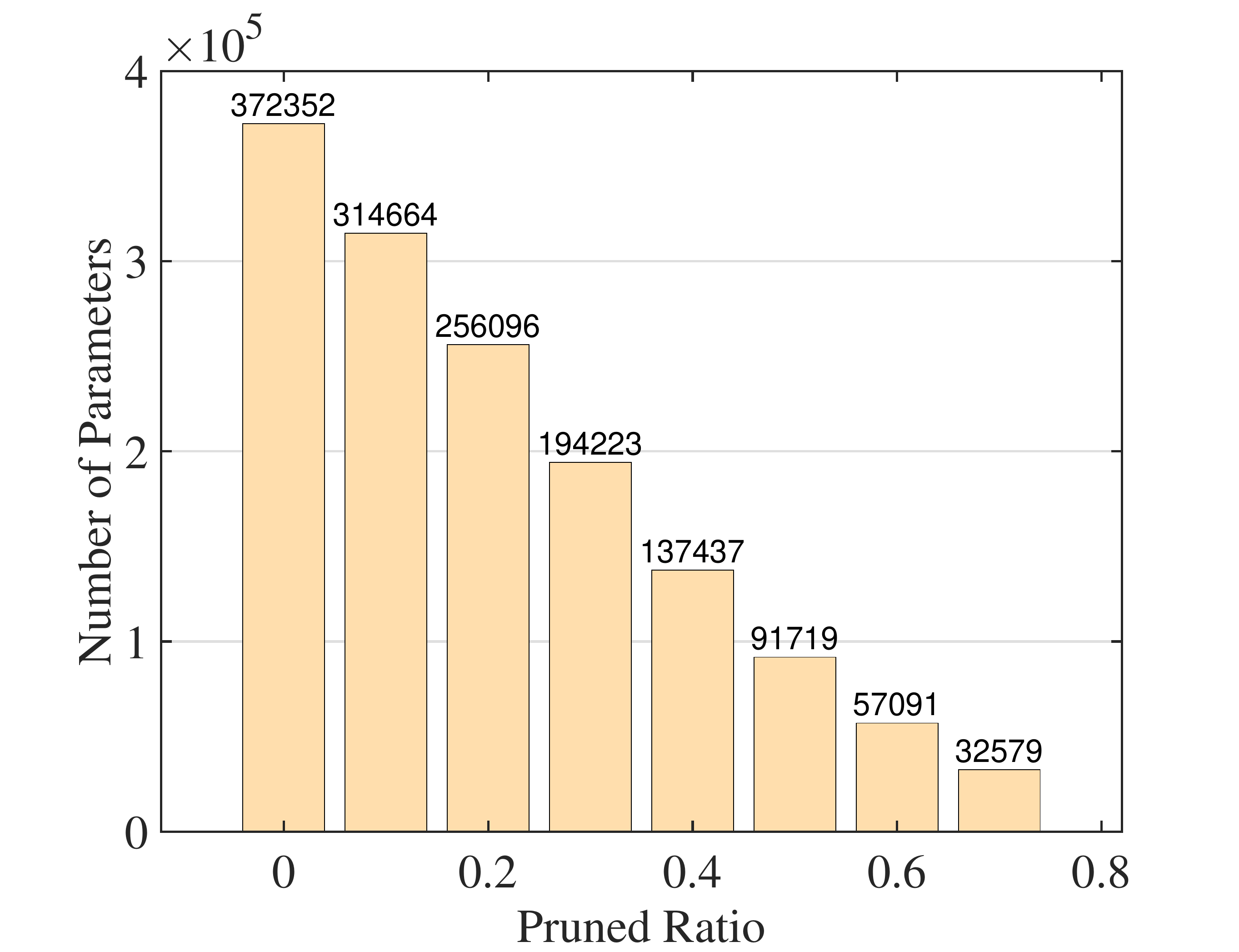}}%%mm_prune
    \subfigure[0.3 THz channel. %\yh{Use the same linewidth as Fig.12(a)}
    ]
    {\includegraphics[width=0.45\textwidth]{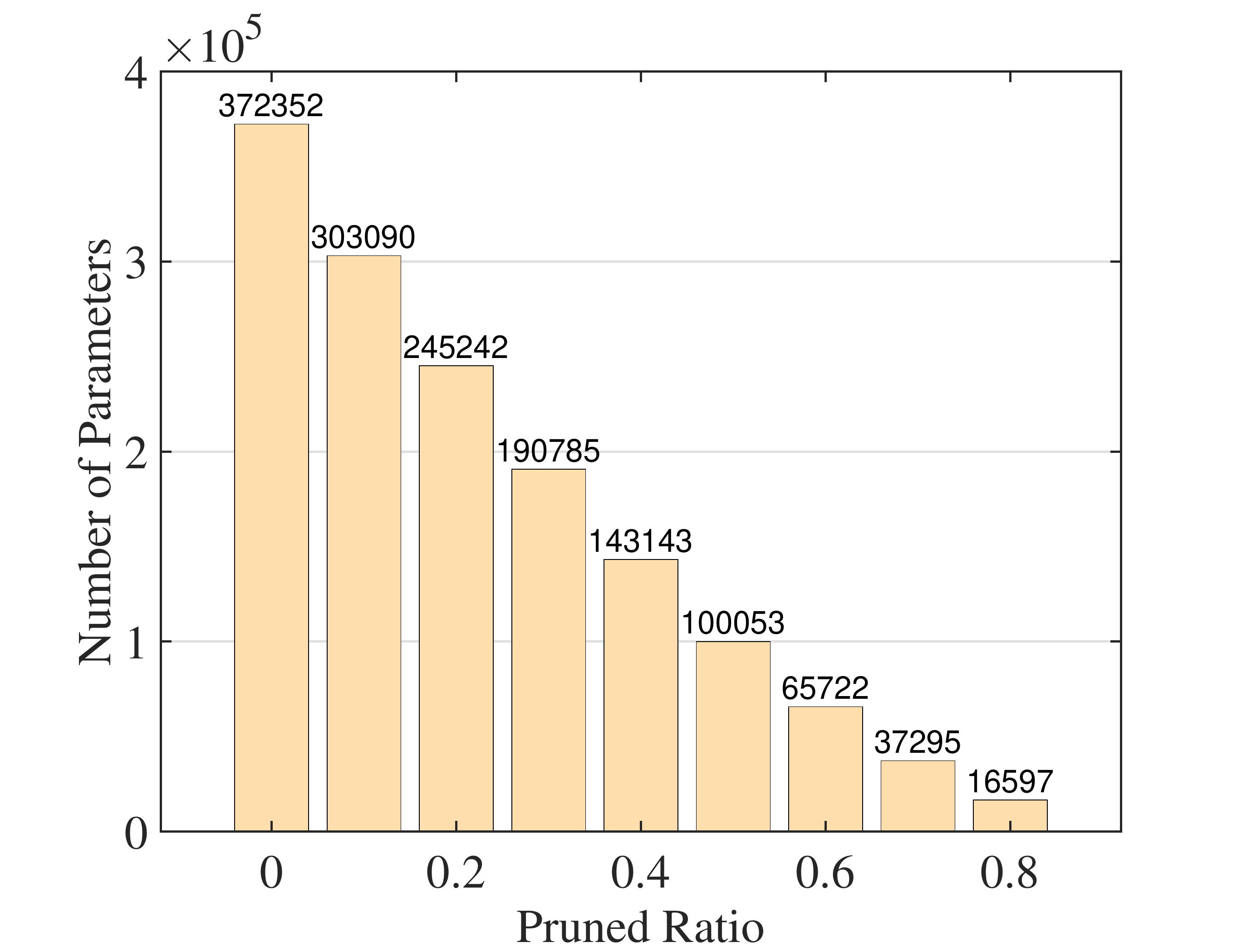}}%%thz_prun
    \caption{Number of parameters in the pruned DCNN with different pruning ratios. }
    \label{fig-size}
    % \vspace{-5mm}
\end{figure}
\begin{figure}[t]
    \centering
    \subfigure[60 GHz channel.]{\includegraphics[width=0.45\textwidth]{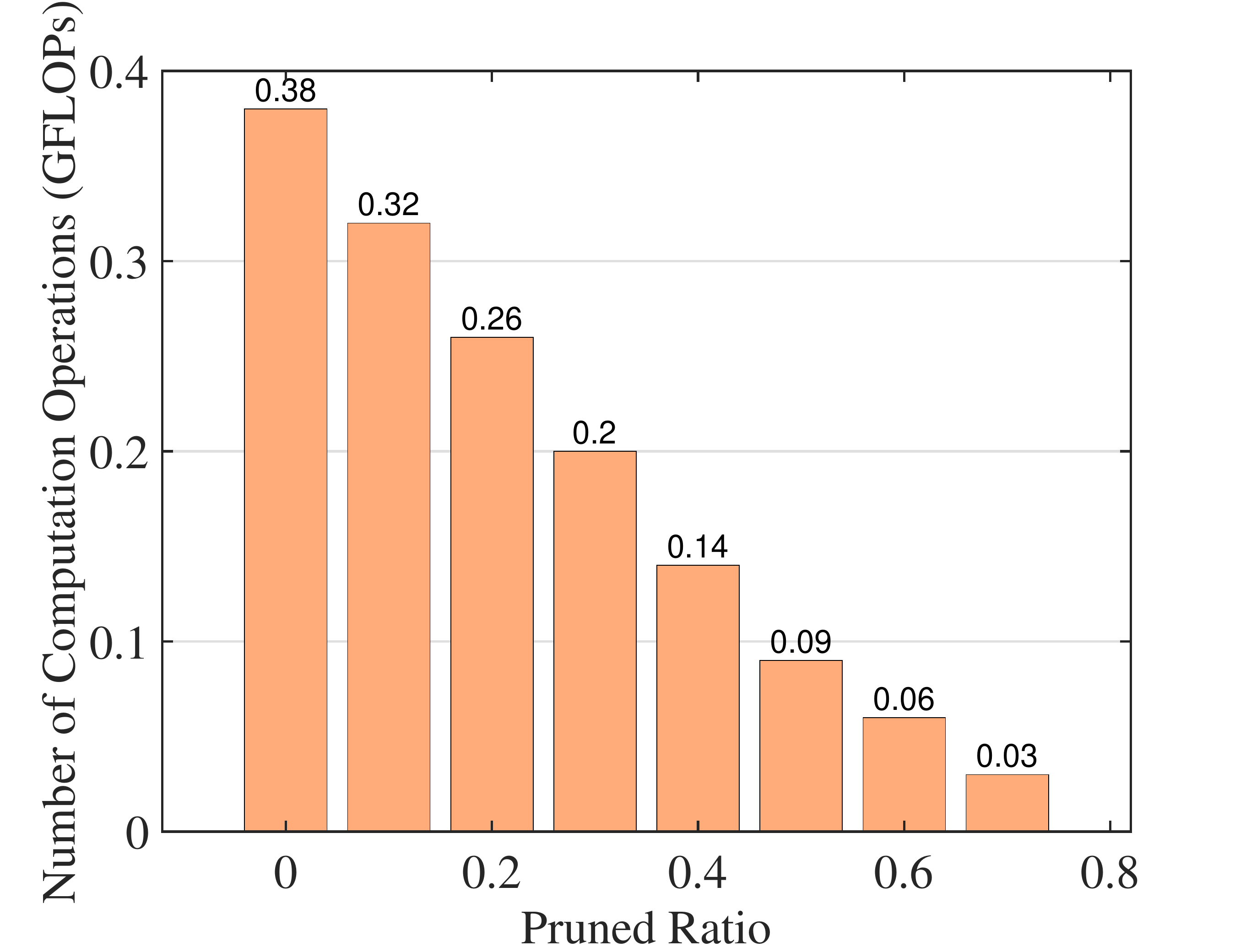}}%%mm_prune
    \subfigure[0.3 THz channel. %\yh{Use the same linewidth as Fig.12(a)}
    ]
    {\includegraphics[width=0.45\textwidth]{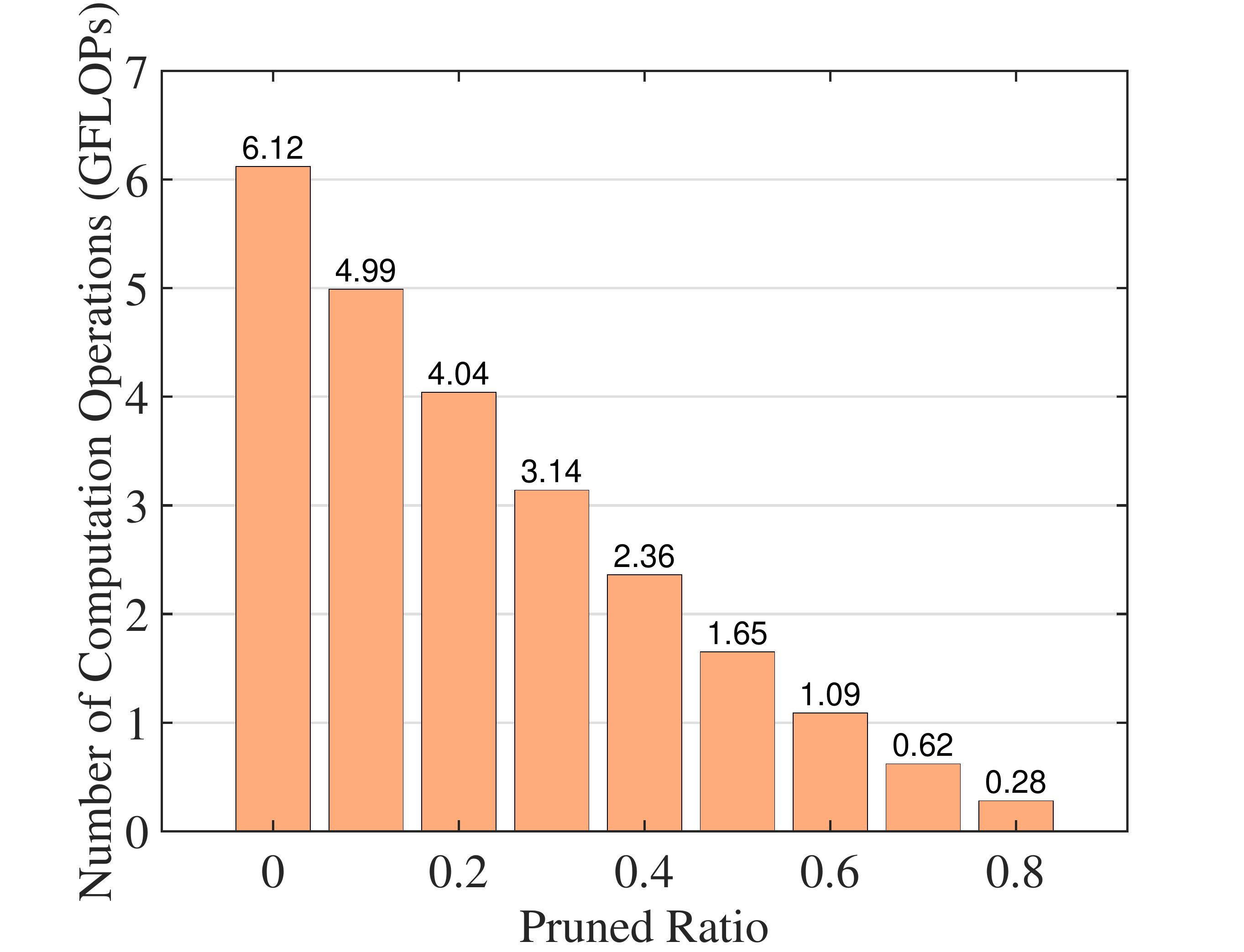}}%%thz_prune
    \caption{Number of computation operations in the pruned DCNN with different pruning ratios.}
    \label{fig-flops}
    % \vspace{-5mm}
\end{figure}

The computational complexities of the aforementioned methods are compared in Table~\ref{tab1}. Particularly, the computational complexity of the OMP algorithm is $\mathcal{O}(SN_tN_r)$, where $S$ denotes the sparsity of the channel, $N_t$ and $N_r$ represent the number of antennas at Tx and Rx, respectively.
Similarly, the computational complexity of AMP algorithm is $\mathcal{O}(T N_tN_r)$, where $T$ denotes the number of iterations in the AMP algorithm. \hdel{LAMP algorithm has the same computation complexity as the AMP algorithm, since LAMP is equivalent to AMP with the learned parameters.}The MMSE estimator has the complexity of $\mathcal{O}(N_t^3N_r^3)$. By contrast, the complexity of the DCNN network is calculated by the dimensions of CV layers, which equals to $\mathcal{O}(N_tN_r\sum_{l=1}^{l_c}F_l^2N_{l-1}N_{l})$, where $l_c$ is the number of CV layers, $F_l$ is the side length of the filters, and $N_l$ is the number of feature maps in the $l^\mathrm{th}$ CV layer. Combining with the tentative estimation module having the complexity of $\mathcal{O}(N_tN_r(N_t+N_r))$, SF-CNN possesses the complexity of $\mathcal{O}(N_tN_r((N_t+N_r)+\sum_{l=1}^{l_c}F_l^2N_{l-1}N_{l}))$. Similarly, the complexity of OMP-DCNN is $\mathcal{O}(N_tN_r(S+\sum_{l=1}^{l_c}F_l^2N_{l-1}N_{l}))$, and the complexity of AMP-DCNN is $\mathcal{O}(N_tN_r(T+\sum_{l=1}^{l_c}F_l^2N_{l-1}N_{l}))$. After pruning the redundant feature maps in the DCNN network, the complexity of the PRINCE is reduced to $\mathcal{O}(N_tN_r(T+\sum_{l=1}^{l_c}F_l^2(N_{l-1}-p_{l-1})(N_{l}-p_l)))$, where $p_l$ describes the reduced number of feature maps for the $l^\mathrm{th}$ CV layer.

The complexities of OMP, AMP and AMP-DCNN increase linearly with the dimension of channel matrix $N_rN_t$, while AMP-DCNN has much better estimation accuracy than \hdel{OMP,}\hzd{the others of them}. On the contrary, the complexity of MMSE estimator grows cubically with the channel dimension, while the estimation accuracy of MMSE is still inferior to AMP-DCNN. SF-CNN, OMP-DCNN and AMP-DCNN have the same complexity for the DL part, since the number of CV layers and feature maps are set the same for the three methods. In terms of the preprocessing module before the DL part, AMP-DCNN has the lowest complexity. Furthermore, the PRINCE is derived from the AMP-DCNN model, and has even lower computation complexity \hzd{than the original AMP-DCNN }thanks to the pruning of redundant feature maps in CV layers.

The complexity of PRINCE decreases as the pruning ratio rises. 
To gain vivid comparison of the complexity between the AMP-DCNN and PRINCE for the mmWave and THz UM-MIMO systems, the number of parameters including the weights and biases for the DCNN part is plotted in Fig~\ref{fig-size}. By pruning the feature maps in the CV layers, the parameters of the corresponding filters and the incoming connections from BN layer are reduced. Specifically, based on the mmWave dataset, the number of parameters is reduced from 372352 to 32579 when seventy percent of feature maps are pruned. Similarly, there is a large reduction in the number of parameters from 373352 to 16597 when the pruning ratio is $80\%$ for THz dataset. Moreover, the number of computation operations of the DCNN part measured by the giga floating point operations (GFLOPs) is depicted in Fig~\ref{fig-flops}. The pruning of the feature maps in CV layers helps reducing the computationally intensive convolution operations, leading that the number of operations decreases from 0.38 GFLOPs to 0.03 GFLOPs for mmWave dataset when the pruning ratio is $70\%$. Moreover, the number of operations for the THz dataset is reduced from 6.12 GFLOPs to 0.28 GFLOPs with a pruning ratio of $80\%$. From Fig.~\ref{fig-size} and Fig.~\ref{fig-flops}, we can observe that the number of parameters and the number of computation operations both shrink quickly with the pruning ratio, which verifies the significant reduction of the complexity of the PRINCE method. 
\section{Conclusion}\label{sec：conclusion}
In this paper, we first proposed a novel AMP-DCNN method, which exploits the benefits of traditional AMP CE method and the DL tool for CE of the mmWave and THz UM-MIMO systems. As a generalized example in deploying DL for CE, AMP is first exploited to obtain the coarse CE result. Then, a designed DCNN refines the results from AMP by further enhancing the estimation accuracy with its powerful learning ability. Based on AMP-DCNN, we further developed a PRINCE method, which reduces the DCNN network size, by truncating the significant feature maps in the CV layers through training with regularization, pruning and refining.

Extensive simulations validate the remarkable estimation accuracy of the proposed AMP-DCNN, whose estimation performance outperforms the benchmark solutions and reaches NMSE of $-12$ dB when SNR is $10$ dB. Moreover, the proposed AMP-DCNN demonstrates robustness in different sizes of training dataset. Furthermore, PRINCE achieves significantly reduced complexity compared to the AMP-DCNN, with negligible performance degradation. With $80\%$ truncated feature maps, the estimation NMSE of PRINCE remains $-10$ dB at SNR = $10$~dB.
%  Furthermore, the PRINCE can achieve a good trade-off between the CE accuracy and low complexity, 

%\hzd{For future work, the pruned DCNN for CE can be further compressed by other techniques, such as  sparsification and quantization. The DL network with tiny size and high accuracy can be obtained to solve the CE problems in practical environments.}% The AMP algorithm is adopted to obtain the coarse estimation of the channel. Then, the DCNN network is designed to refine the output of AMP to yield the final estimation result with high accuracy.

%\yh{further revise：Please unify the format of references}
\bibliographystyle{IEEEtran}
%\bibliography{main}

\end{document}